\documentclass{article}
\usepackage[dvips]{graphicx}
\usepackage{amsmath}
\usepackage{amsfonts}
\usepackage{amssymb}

\begin{document}
\title{Crowd-Anticrowd Theory of Multi-Agent Minority Games}
\author{Michael L. Hart and Neil F. Johnson\thanks{n.johnson@physics.ox.ac.uk}\\
Physics Department, Oxford University,  Oxford, OX1 3PU, UK}

\maketitle

\abstract{We present a formal treatment of the Crowd-Anticrowd theory of Minority
Games played by a population of competing agents. This theory is built around
a description of the crowding which arises within the game's strategy space. Earlier
works have shown that this theory provides a simple, yet quantitatively accurate
explanation of the time-averaged behavior of these multi-agent
games. We also discuss the extent to which the Crowd-Anticrowd approach provides a
useful tool for analyzing a wider class of Complex Systems.}

\newpage

\section{Introduction}
The now-famous Minority Game (MG) was introduced by Challet and Zhang \cite{origMG}
as a means of simplifying Brian Arthur's original El Farol Bar
Problem
\cite{farol,uselfarol}. The fascination with the MG among researchers from physics,
computer science, biology and economics, has led to the appearance of more than
one hundred MG-related papers within the past few
years (see, for example, Refs.
\cite{RSS,ChalletContTime,ChalletPhasTran,ChalletSpin,ChalletStyle,
marsili2,marsiliThermo,memory,comment,SavitLet,SavitPhysA,
SavitPhysAb,CAT1,CAT2,alloy,generalTherm,CAA,MarketModel,
THMG1,THMG2,paul,prl,heimel,heimel2,cavagna,SherThermLett,
SherMarkMod}). A detailed list, with informative comments by Damien Challet, is
available from Ref.
\cite{webpage}. The game itself concerns a population of $N$\ heterogeneous agents
with limited capabilities and information, who repeatedly compete to be in the
minority group. The agents (e.g. people, cells, data-packets) are adaptive, but only
have access to global information about the game's progress. 

The numerical results obtained by Savit and co-workers \cite{SavitLet} prompted
widespread interest in the Minority Game. These results showed that the time-averaged
fluctuations in the game's output -- or equivalently, the time-averaged fluctuations
in wastage of the underlying global resource -- vary in a highly non-linear fashion 
as a function of the memory
$m$ of the agents. Reference \cite{CAT1} provided the first explanation of this
`Savit curve'. In particular it was shown that the formation of, and competition
between, crowds and anticrowds can explain both qualitatively and quantitatively the
Savit curve for the basic MG
\cite{CAT1,CAT2}. In addition, the Crowd-Anticrowd approach has been shown to
describe the fluctuations within more general MGs: for example, an MG  comprising
a population of agents with different memories
$m$ \cite{alloy,dublin}, an MG with stochastic strategy-use \cite{generalTherm}, and
an MG with a mixed population of stochastic and non-stochastic strategy-use
\cite{CAA}. The task of fully understanding the MG is seen as an important stepping
stone toward the development of a `Theory of Complex Systems'. Many systems have been
proposed as real-world examples of the MG, in fields as diverse as biology,
computing science, sociology, economics and finance. (Note however that the basic MG
lacks the crucial feature present in financial markets whereby agents can sit out of
the game and hence do nothing at any given timestep. Including this effect yields the
Grand Canonical Minority Game (GCMG) as discussed in Refs.
\cite{dublin,MarketModel}, which can indeed show financial market-like behavior).

In addition to the Crowd-Anticrowd approach, there have been various more recent
attempts to develop a quantitative theory which describes the fluctuations as a
function of 
$m$ in the MG \cite{ChalletContTime,ChalletPhasTran,ChalletSpin,ChalletStyle,
marsili2,memory,heimel,heimel2,SherThermLett,
SherMarkMod}. Although elegant and
sophisticated, such theories have not been able to reproduce the numerical results
of Ref. \cite{SavitLet} over the full range of $m$. We believe that what is
missing from such theories is an accurate description of the correlations between
agents' strategies: these correlations produce a highly-correlated form of decision
noise which cannot then be easily averaged over or added in. By contrast these strong
inter-strategy correlations take center-stage in the Crowd-Anticrowd theory.

The Crowd-Anticrowd approach is not limited to MG-like games. It is built
around the effects  of crowding (i.e. correlations) in strategy-space, rather than
the precise rules of the game itself, and only makes fairly modest assumptions about
a game's dynamical behavior. Specifically, its validity depends on
the histories of the game (see Sec. 2) being visited reasonably frequently. As a
result, it is likely that it is applicable to other forms of multi-agent game.
To the extent that a given Complex System's dynamics mimic such a multi-agent game,
it is likely that the Crowd-Anticrowd approach will be applicable. This would be a
welcome development, given the lack of general concepts in the field of Complex
Systems as a whole. It is therefore important to lay out the theoretical framework
underlying the theory, before proceeding with any broader investigations. 

Despite the potential usefulness of the Crowd-Anticrowd theory, the formal
derivation has not yet been presented in the literature. This observation motivates
the present paper. The layout of the paper is as follows. For completeness, Section
2 provides a description of the basic MG. Section 3 presents a derivation of the
Crowd-Anticrowd theory, while Section 4 looks at various limiting cases. The
conclusion is given in Section 5.

\section{Description of the Minority Game}
Figure 1 summarizes the Minority Game (MG). At timestep $t$, each agent (e.g. a bar
customer, a commuter, or a market trader) decides whether to enter a game where the
choices are option $1$ (e.g. attend the bar, take route A, or buy) and option $0$
(e.g. go home, take route B, or sell). A total of $n_{0}(t)$\ agents choose $0$ while
$n_{1}(t)$\ choose
$1$. We can define an excess demand or net
`attendance' as
\begin{equation} A(t)=n_{1}(t)-n_{0}(t).  \label{vol}
\end{equation}
The only global information available to the agents is a common
memory of the recent history, i.e. the most recent $m$ winning decisions/outcomes.
For $m=2$, these will have the form $00$, $01$, $10$ or $11$. Hence at each
timestep, the recent history constitutes a particular bit-string. For general $m$,
there will be $P=2^{m}$ possible history bit-strings. These history bit-strings can
alternatively be represented in decimal form: $\mu =\{0,1,...,P-1\}$ where $\mu =0$\
corresponds to $00$, $\mu =1$\ corresponds to 01 etc.  A strategy consists of a
response, $0$ or $1$, to each possible history bit-string. Hence there are $2^{P}$
possible strategies. For
$m=2$ for example, there are therefore $16$ possible strategies. Each agent randomly
picks $s$ strategies at the outset of the game. The agents update the scores of
their strategies after each timestep with $+1$ (or $-1$) as the pay-off for choosing
the winning (or losing) action. Agents play their highest scoring strategy. If an
agent holds two or more strategies that are tied for the position of `highest scoring
strategy' then the agent will use a fair random process (e.g. a coin-toss) to decide
which of these strategies to use for that turn of the game. 

A feature of the MG which has attracted much interest, is the standard deviation (or
`volatility') of the number of agents choosing a particular option over time, i.e.
the standard deviation of
$n_{0}(t)$ or
$n_{1}(t)$ \cite{SavitLet}. The time-averages of $n_{0}(t)$ and $n_{1}(t)$
are both approximately $N/2$ since neither 0 nor 1 is preferred as an outcome a
priori.  Only those agents who choose the correct minority
option at a given turn get positively rewarded, hence the standard deviation of
$n_{0}(t)$ or
$n_{1}(t)$ (and $A(t)$) indicates the level of wastage in the system. If the
standard deviation of
$n_{0}(t)$ is small then a relatively large proportion of agents are
getting rewarded positively at each timestep, and hence the level of wastage 
$\left\vert n_{0}(t)-N/2\right\vert $ is small. Conversely, if the
standard deviation of $n_{0}(t)$ is large then a relatively small proportion of
agents are getting rewarded positively at each timestep, and hence the level of
wastage 
$\left\vert n_{0}(t)-N/2\right\vert $ is large. 

Figure 2 shows the `Savit curve'. This plots the standard deviation (i.e. volatility)
of agents choosing a particular option,  as a function of memory size
$m$. The numerical values are shown as small circles, and were obtained from
individual simulation runs.  The data for each run is
collected once initial transient effects have settled down. The volatilities from
different runs differ because of different initial strategy allocations among the
agents, and different outcomes from the coin-tosses used to break ties in strategy
scores. The dashed line represents the volatility one would get if all the agents
used the toss of a coin to decide which option to choose at every timestep. To see
how this is calculated, consider the case of $N$ independent agents each deciding
which option to choose by the toss of a coin. Each agent therefore provides a
random-walk process in terms of increasing or decreasing
$A(t)$ by $1$. Assume for the moment that these coin-tosses are uncorrelated. Then
the total variance of this random-walk in the attendance $A(t)$, is given by the sum
of the individual variances produced by each of the $N$ agents. If the agent decides
$1$, then he contributes $1$ to the attendance $A(t)$. [The random-walk
\textquotedblleft step-size\textquotedblright\ is $d=1$]. If, by contrast, the agent
decides $0$, then he contributes $-1$ to the attendance $A(t)$. The agent chooses
$1$ with probability
$p=1/2$, and $0$ with probability $q=1/2$. The variance contributed to $\sigma ^{2}$
by each agent is therefore given by the random-walk result $4pqd^{2}=1$. Summing
over all $N$ agents, the total variance in the excess demand is given by
$4Npqd^{2}=N$. The variance
$\sigma _{1}^{2}$\ of $n_{1}(t)$\ can be obtained from the variance $\sigma ^{2}$ of
$A(t)$ by dividing by a factor $4$ [since $2\sigma _{1}=\sigma$]. Hence the
variance $\sigma _{1}^{2}$\ of $n_{1}(t)$ is given by $N/4$.  The corresponding
standard deviation $\sigma_1={\sqrt N}/2$ is the dashed `coin-toss' line in the plot
(i.e.
$\sigma_1\approx 5.0$ for $N=101$). The solid lines in Figure 2 correspond to
analytic Crowd-Anticrowd results calculated in various regimes: these are
discussed later in the paper.

\subsection{Strategy Space}

A strategy is defined as a set of instructions to describe what an agent should
do in any given situation, i.e. given any particular history $\mu$ the
strategy decides what the agent should do. The strategy space is the set of
strategies from which agents are allocated their strategies at the beginning
of the game. Any strategy in the strategy space can in principle be present in the
game: however if a strategy is not in the set which is initially allocated, then it
can never appear in that particular run of the MG game. Figure 3 shows an
$m=2$ strategy space together with some example strategies. The strategy space shown
is known as the `Full Strategy Space', FSS, and contains all possible permutations of
the binary options
$0$ and $1$ for each history. As such there are $2^{2^{m}}$ strategies in this
space. The $2^{m}$ dimensional hypercube shows all 
$2^{2^{m}}$ strategies from the full strategy space at its vertices. It is clear
that the agents playing the game are limited by the set of strategies that they
are allocated at the start of the game. (This initial strategy-allocation is random).
Of course there are many more strategies that can be thought of, but which aren't
present within the FSS. For example, the simple strategies of persistence and
anti-persistence are not present in the FSS. The advantage however of using the FSS
in the MG is that the strategies form a complete set and as such display no bias
towards any option given a history. There are also practical reasons for using a
binary strategy space. Consider a game with a memory size of $m=8$, i.e. the agents
can look at the last $8$ outcomes of the game to decide what to do next. Even with
binary options for a given history there are a large number of
strategies -- in particular there are $2^{2^{m}}=1.16\times 10^{77}$.
To include any additional strategies like
persistence and anti-persistence would mean opening up the strategy space, hence
losing the simplicity of the MG and returning to the complexity of Arthur's original
Bar Problem
\cite{farol,uselfarol}.

\subsection{Reduced Strategy Space (RSS)}

It can be observed from the FSS, that one can choose a subset of strategies
\cite{RSS} such that any pair within this subset has one of the following
characteristics:

\begin{itemize}
\item anti-correlated, e.g. $0000$ and $1111$. For example, any two agents using the
($m=2$) strategies $0000$ and $1111$ respectively, would take the opposite action
irrespective of the sequence of previous outcomes and hence the history. Hence one
agent will always do the opposite of the other agent. For example, if one agent buys
at a given timestep, the other agent will sell. Their net effect on the attendance
$A(t)$ therefore cancels out at each timestep. Hence they will not contribute to
fluctuations in $A(t)$. In short they do not contribute to the volatility. This is a
crucial observation for understanding the behaviour of the volatility in this
system, as we discuss later on.

\item uncorrelated, e.g. $0000$ and $0011$. For example, any two agents using the 
strategies 
$0000$ and $0011$ respectively, would take the opposite action for two of the four
histories, while they would take the same action for the remaining two histories.
Assuming that the $m=2$ histories occur equally often, the actions of the two agents
will be uncorrelated on average.
\end{itemize}

\noindent A convenient measure of the distance (i.e. closeness) of any two strategies
is the Hamming distance which is defined by the number of bits that need to be
changed in going from one strategy to another. For example, the Hamming distance
between $0000$ and $1111$ is $4$, while the Hamming distance between $0000$ and
$0011$ is just $2$. Although there are $2^{P}\equiv 2^{2^{m=2}}\equiv 16$\ strategies
in the
$m=2 $ strategy space, it can be seen that one can choose subsets such that any
strategy-pair within this subset is either anti-correlated or
uncorrelated. Consider, for example, the two groups
\begin{equation*} U_{m=2}\equiv \{0000,1100,1010,0110\}
\end{equation*} 
and
\begin{equation*}
\overline{U_{m=2}}\equiv \{1111,0011,0101,1001\}.
\end{equation*} 
Any two strategies within $U_{m=2}$\ are uncorrelated since they
have a Hamming distance of $2$. Likewise any two strategies within
$\overline{U_{m=2}}$\ are uncorrelated since they have a relative Hamming distance of
$2$. However, each strategy in $U_{m=2}$ has an anti-correlated strategy in
$\overline{U_{m=2}}$: for example, $0000$ is anti-correlated to $1111$, $1100$ is
anti-correlated to $0011$ etc. This subset of strategies comprising $U_{m=2}$\ and
$\overline{U_{m=2}}$ , forms a Reduced Strategy Space (RSS) \cite{RSS}. Since it
contains the essential correlations of the Full Strategy Space (FSS), it turns out
that running the simulation within the RSS reproduces the main features of the full
game obtained using the FSS \cite{RSS}. The RSS has a smaller number of strategies
$2.2^{m}=2P\equiv 2^{m+1}$ than the FSS which has $2^{P}=2^{2^{m}}$. For $m=2$,
there are $8$ strategies in the RSS, compared to $16$ in the FSS, whilst for $m=8$
there are
$1.16\times 10^{77}$ strategies in the FSS, but only $512$\ strategies in the RSS.
We note that the choice of the RSS is not unique, i.e. within a given FSS there are
many possible choices for a RSS. In particular, it is possible to create
$2^{2^{m}}/2^{m+1}$ distinct reduced strategy spaces from the FSS. To summarize, the
RSS provides a minimal set of strategies which `span' the FSS and are hence
representative of its full structure.

\subsection{History Space: de Bruijn graph}

The history $\mu $ of recent outcomes changes in time, i.e. it is a dynamical
variable. Interestingly the dynamics of this history can be represented on a
directed graph (a so-called digraph). The particular form of directed graph is
called a de Bruijn graph. Figure 4 shows some examples of the de Bruijn graph for
$m=1,2,$ and $3$. The probability that the outcome at time $t+1$ will be a $1$ (or
$0$) depends on the state at time $t$. Hence it will depend on the previous $m$
outcomes, i.e. it depends on the particular state of the history bit-string. The
dependence on earlier timesteps means that the game is non-Markovian. We note
that modifying the game to a finite time-horizon for scoring strategies, allows the
resulting game to be viewed as a high-dimensional Markov process \cite{THMG1,THMG2}.
However, here we focus on the basic MG where scores are kept from the beginning of
the game.

\subsection{Initial conditions for the MG}

The dynamics for a particular run of the game depend upon the strategies that the
agents are initially assigned and the random process used to decide tie-breaks. The
particular dynamics which emerge also depend upon the initial score-vector of the
strategies and initial history used to seed the start of the game. If the initial
strategy score-vector is not `typical', then a bias can be introduced into the game
which never disappears. In short, the system never recovers from this bias. It will
be assumed that no such initial bias exists. In practice this is achieved, for
example, by setting all the initial scores to zero. The initial choice of history is
not considered to be an important effect. It is assumed that any transient effects
resulting from the particular history seed will have disappeared, i.e. the initial
history seed does not introduce any long-term bias. A typical game is left to run
for over $10,000$ timesteps such that any transients are washed out of the system.

The initial strategy allocation among agents can be described in terms of a tensor
$\Omega
$ \cite{paul}. This tensor $\Omega $ describes the distribution of strategies among
the $N$ individual agents and hence the particular \emph{quenched disorder} of the
system. The dimension of $\Omega $\ is given by the number of strategies $s$ that
each agent holds. For example, for $s=3$ the element $\Omega _{i,j,k}$ gives the
number of agents assigned strategy $i$, then strategy $j$, and then strategy 
$k$, in that order. Hence
\begin{equation}
\sum_{i,j,k,...}^{D}\Omega _{i,j,k,...}=N,
\end{equation}
where the value of $D$\ represents the number of distinct strategies
that exist within the strategy space chosen. $D=2^{2^{m}}$ in the FSS, whilst
$D=2.2^{m}$ in the RSS. Figure 5  shows an example distribution $\Omega$ for $N=101$
agents in the case of $m=2$, and $s=2$ in the reduced strategy space RSS. 
We note that while $\Omega$ is not symmetric, it could be made so since the MG does
not distinguish between the order in which the two strategies are picked.

\section{Derivation of Crowd-Anticrowd theory}

\subsection{Qualitative explanation} 
We start by providing a qualitative explanation of the `Savit curve' in Figure 2
(small circles). Our explanation is based on the RSS, but we note that Figure 2 is
practically identical for both RSS and FSS. First we introduce the notion of {\em
crowd} and {\em anticrowd}. Consider a group of agents whose highest-scoring
strategy is $R$ over
$\delta t$ timesteps. This group constitutes a {\em crowd} since they all use this
same strategy
$R$ and hence act in the same way for those $\delta t$ timesteps {\em
regardless} of the particular history bit-string at each timestep. 
Next consider the group of agents whose highest-scoring strategy over the same $\delta
t$ timesteps is the anti-correlated partner to $R$, which we refer to as $\overline R$.
This group constitutes an {\em anticrowd} with respect to the crowd using strategy
$R$, since they will act in the opposite way to the first crowd at each timestep {\em
regardless} of the particular history bit-string. 

Figure 6 indicates the
competition between crowd-anticrowd sizes which arises as a function of $m$. The
curve in the upper graph shows the volatility from Figure 2 averaged over runs, as a
function of
$m$. The low-$m$ phase, characterized by a decrease in
$\sigma $\ as $m$ increases, is called the 
crowded phase since the number of strategies in the RSS $2^{m+1}$ is small
compared to the number of agents $N$. The high-$m$ phase, characterized by a slow
increase in $\sigma $ towards a limiting value as $m$ increases, is called the
dilute phase since the number of strategies in the RSS $2^{m+1}$ is now large
compared to the number of agents $N$. 

In the crowded phase, i.e. at small $m$,
there will at any one time be a large number of agents who are using a given strategy
$R$ and so will flood into the market as a crowd. Although the crowd may
maintain strategy $R$ for several timesteps, the resulting outcome will tend to flip
between  0 and 1 in response to the changing  history bit-string (consider, for
example, strategy 0011 which produces equal numbers of 0's and 1's as the game
cycles through history bit-strings). This gives rise to a high volatility.
Eventually, strategy $R$ will lose enough points to be deemed a bad strategy by many
of the crowd members. Hence the crowd using a given strategy $R$ is continually
changing its membership and size. We emphasize
that the crowding effect we are discussing at any given timestep will occur {\em
regardless} of the actual history bit-string at that timestep, since it is a
crowding effect in strategy space rather than merely a crowding in terms of eventual
action. In other words, we are not just making the trivial statement that a crowd
comprises agents who act in the same way at a given timestep --  instead we are
saying that a crowd is characterized by agents who all believe that a given strategy
is the best, and will therefore take the same action regardless of the history
bit-string at a particular timestep. 

As the memory
$m$ of the agents increases, the number of agents using high-scoring strategies 
tends to decrease simply because many may not now possess such
strategies. This implies a decrease in
the crowd-sizes, but an increase in the number of crowds reflecting an increase in
the number of strategies available. There will also be groups of agents who are
forced to use the anti-correlated (i.e. low-scoring) strategies. The actions of
these  anticrowds will cancel out the action of the crowds. This argument applies to
all pairs of anti-correlated strategies, i.e. the crowd-anticrowd pair corresponding
to the best/worst strategy pair, the crowd-anticrowd pair corresponding to the
second-best/second-worst strategy pair, etc. As $m$ increases, the crowd-sizes
decrease while the anticrowd sizes increase: this  yields an increase in the
crowd-anticrowd  cancellation effect and hence a reduction in the size of
the volatility. Eventually in the dilute phase of very large memory
$m$, it is very unlikely that any agents will use (or even hold) the same strategies.
It is also unlikely that any two agents will be using anti-correlated strategies.
Hence the crowd-anticrowd cancellation is reduced, thereby increasing the
volatility. Because of the lack of correlated groups (i.e. all crowds and anticrowds
are of size 1 or 0) the volatility can now be modelled by assuming that the
population comprises independent, coin-tossing agents.

We now add some analytics to this verbal description. 
Consider the crowd of agents
$n_{R}$ using a particular strategy $R$ at a particular moment during the game, 
and the anticrowd of $n_{\overline{R}}$ agents who are using the
anticorrelated strategy to $R$. Over the timescale during which the two opposing
strategies are being played, the fluctuations are determined only by the net
crowd-size
$n_{R}^{eff}=n_{R}-n_{\overline{R}}$ which constitutes the net step-size of the
crowd-anticrowd pair in a random-walk model for the attendance. The net
contribution to the volatility by this crowd-anticrowd pair is therefore 
$pqd^{2}=[n_{R}^{eff}]^{2}/4$. This simple expression enables us to discuss the
order of magnitude of the volatility within the three main regions of the curve seen
in Figure 2:

\begin{itemize}
\item $2^{m+1}<<N.s$ : suppose strategy $R^{\ast }$\ is the highest scoring at a
particular moment. The anti-correlated strategy $\overline{R^{\ast }}$\ is therefore
the lowest scoring at that same moment. In the limit of small $m$, the size of the
strategy space is small. Each agent hence carries a considerable fraction of all
possible strategies. Therefore, even if an agent picks $\overline{R^{\ast }}$ among
his $s$ strategies, he is also likely to have a high scoring strategy. Therefore,
many agents will choose to use either $R^{\ast }$ itself (if they hold it) or a
similar one. In this regime the
crowd associated with the highest-scoring strategies will dominate. Therefore
$n_{R}\thicksim N\delta _{RR^{\ast }}$\ and hence $n_{R}^{eff}\thicksim N\delta
_{RR^{\ast }}$. This implies that the variance varies as $N^{2}/4$, i.e. the
volatility is $\sigma\sim N/2=50$ for $N=101$.

\item $2^{m+1}>>N.s$ : in the limit of large m, the strategy space is very large and
agents will have a low chance of holding the same strategy. Even if an agent has
several low-scoring strategies, the probability of his best strategy being strictly
anticorrelated to another agent's best strategy (hence forming a crowd-anticrowd
pair) is small. All the crowds and anticrowds will tend to be of size $1$ or $0$,
hence there are of order $N$ contributions to the variance, each with effective
step-size
$n_{R}^{eff}=1$. This implies that the variance varies as
$N[n_{R}^{eff}]^{2}/4=N/4$, i.e. the volatility is $\sigma\sim \sqrt{N}/2=5$ for
$N=101$.

\item $2^{m+1}\thicksim N.s$ : in the intermediate $m$ region where the numerical
minimum exists, the size of the strategy space is relatively large. Hence some
agents may get stuck with $s$ strategies which are all low scoring. They hence form
anticrowds. Considering the extreme case where the crowd and anticrowd are of
similar size, this gives $n_{R}^{eff}\sim 0$\ and hence the volatility
$\sigma\rightarrow 0$. For a fixed value of $m$, this regime of small volatility
will arise for small $s$ since, in this case, the number of strategies available to
each agent is small - hence some of the agents may indeed be forced to use a strategy
which is little better than the poorly-performing $\overline{R^{\ast }}$. In other
words, the cancellation effect of the crowd and anticrowd becomes most effective in
this intermediate region for small
$s$. Increasing $s$ will make this minimum less marked since it will reduce the
number of agents forced to use the anti-correlated strategy, hence reducing the
crowd-anticrowd cancellation effect in the volatility (as shown later in Figure 11).
\end{itemize}

\noindent We note that this Crowd-Anticrowd concept, and the subsequent detailed
theory presented below, does not use any knowledge of the specific history bit-string
at a given time-step. Hence the theory does not depend on the detailed dynamics
of the history bit-string, as long as the dynamics is such that
all histories (i.e. nodes in the de Bruijn graph) are visited frequently thereby
guaranteeing that the measure of correlation for strategy pairs will have meaning.
Hence the crowd-anticrowd theory would predict the same volatility for the MG
regardless of whether the real history was used at each timestep, or whether it was
replaced by a random bit-string. This is indeed what has been found in numerical
studies of the volatility by Cavagna
\cite{cavagna}, thereby establishing the usefulness of the Crowd-Anticrowd approach.

\subsection{Proof of concept}

Having given a qualitative understanding of the underlying physics, we will now provide
a numerical proof-of-concept of the Crowd-Anticrowd approach  before giving a formal
derivation of the theory itself.  Consider a given realization of the quenched
disorder
$\Omega
$ and a timestep $t$ in a given run for this given $\Omega $. There is a current
score-vector $\underline{S}[t]$ and a current history $\mu \lbrack t]$\ which define
the state of the game. The attendance $A(t)=A\left[ \underline{S}[t],\mu \lbrack
t]\right]$ is given by Equation 1. 
The volatility for a given run corresponds
to a \emph{time-average}: it is the standard deviation of the number of agents
making a given decision (e.g. $1$) for a given realization of the quenched disorder
$\Omega
$ and a given set of initial conditions. The volatility will eventually be averaged
over many runs, and hence will be averaged over all realizations of the quenched
disorder
$\Omega $ \emph{and} all sets of initial conditions. However, consider first a
\emph{given} realization of the quenched disorder $\Omega$. We will assume that the
quantities of interest (i.e. the mean and standard deviation of the attendance
$A(t)$, and hence also $n_{1}(t)$ and $n_{0}(t)$) self-average for a given
realization of the quenched disorder $\Omega $. In other words, it is assumed that
the average over time is equivalent to an average over initial conditions for
the given realization of the quenched disorder $\Omega $. We have checked using
numerical simulations that this assumption is valid.
The attendance in Equation 1 can be rewritten by summing over the RSS as
follows:
\begin{equation} A\left[ \underline{S}[t],\mu \lbrack t]\right]
=n_{1}(t)-n_{0}(t)\equiv \sum_{R=1}^{2P}a_{R}^{\mu
\lbrack t]}n_{R}^{\underline{S}[t]},
\end{equation}
where $P=2^{m}$\ and the quantity $a_{R}^{\mu \lbrack t]}$ is the
response of strategy $R$ to the history bit-string $\mu $\ at time $t$. Option $1$
corresponds to $a_{R}^{\mu \lbrack t]}=1$ while option $0$ corresponds to
$a_{R}^{\mu \lbrack t]}=-1$. The quantity $n_{R}^{\underline{S}[t]}$ is the number
of agents using strategy $R$ at time $t$. [The superscript $\underline{S}[t]$ is a
reminder that this number of agents will depend on the strategy score at time $t$].
The calculation of the average attendance will now be shown, where the average is
over time for a given realization of the quenched disorder $\Omega $. $\left\langle
X(t)\right\rangle _{t}$ is defined as a time-average over the variable $X(t)$ for a
given $\Omega $. By assuming the self-averaging property for a given $\Omega $, the
system is essentially assumed to be ergodic: hence we can assume that all histories
will be visited with similar frequency in a given run. Hence
\begin{eqnarray}
\left\langle A\left[ \underline{S}[t],\mu \lbrack t]\right] \right\rangle _{t}
&=&\sum_{R=1}^{2P}\left\langle a_{R}^{\mu \lbrack
t]}n_{R}^{\underline{S}[t]}\right\rangle _{t} \\ &=&\sum_{R=1}^{2P}\left\langle
a_{R}^{\mu \lbrack t]}\right\rangle _{t}\left\langle
n_{R}^{\underline{S}[t]}\right\rangle _{t}  \notag
\\ &=&\sum_{R=1}^{2P}\left( \frac{1}{P}\sum_{\mu =0}^{P-1}a_{R}^{\mu \lbrack
t]}\right) \left\langle n_{R}^{\underline{S}[t]}\right\rangle _{t}  \notag \\
&=&\sum_{R=1}^{2P}0.\left\langle n_{R}^{\underline{S}[t]}\right\rangle _{t} 
\notag \\ &=&0.  \notag
\end{eqnarray}
Notice that the averaging is performed over all the histories,
because of the ergodic assumption. The interest is in the \emph{fluctuations} of
$A(t)$ about this average value. Hence the volatility of $A(t)$ is considered. The
variance (or volatility squared) is given by
\begin{eqnarray}
\sigma _{\Omega }^{2} &=&\left\langle A\left[ \underline{S}[t],\mu \lbrack t]
\right] ^{2}\right\rangle _{t}-\left\langle A\left[ \underline{S}[t],\mu
\lbrack t]\right] \right\rangle _{t}^{2}  \label{vol2} \\ &=&\left\langle A\left[
\underline{S}[t],\mu \lbrack t]\right] ^{2}\right\rangle _{t}  \notag \\
&=&\sum_{R,R^{\prime }=1}^{2P}\left\langle a_{R}^{\mu \lbrack
t]}n_{R}^{\underline{S}[t]}a_{R^{\prime }}^{\mu \lbrack t]}n_{R^{\prime
}}^{\underline{S}[t]}\right\rangle _{t}.  \notag
\end{eqnarray}
This double sum is now broken into three parts: \underline{$a_{R}$}$.
\underline{a_{R^{\prime }}}=P$\ (correlated), \underline{$a_{R}$}$.
\underline{a_{R^{\prime }}}=-P$\ (anti-correlated), and \underline{$
a_{R}$}$.\underline{a_{R^{\prime }}}=0$ (uncorrelated). While this cannot
be done in the FSS, the decomposition is exact in the RSS. Hence we have
\begin{eqnarray}
\sigma _{\Omega }^{2} &=&\sum_{R=1}^{2P}\left\langle \left( a_{R}^{\mu
\lbrack t]}\right) ^{2}\left( n_{R}^{\underline{S}[t]}\right) ^{2}\right\rangle
_{t}+\sum_{R=1}^{2P}\left\langle a_{R}^{\mu \lbrack t]}a_{
\overline{R}}^{\mu \lbrack t]}n_{R}^{\underline{S}[t]}n_{\overline{R}}^{
\underline{S}[t]}\right\rangle _{t}+
\sum_{R\neq R^{\prime }\neq \overline{R}}^{2P}\left\langle a_{R}^{\mu \lbrack
t]}a_{R^{\prime }}^{\mu \lbrack t]}n_{R}^{\underline{S}[t]}n_{R^{\prime
}}^{\underline{S}[t]}\right\rangle _{t}  \notag \\ &=&\sum_{R=1}^{2P}\left\langle
\left( n_{R}^{\underline{S}[t]}\right)
^{2}-n_{R}^{\underline{S}[t]}n_{\overline{R}}^{\underline{S}[t]}\right\rangle
_{t}+\sum_{R\neq R^{\prime }\neq \overline{R}}^{2P}\left\langle a_{R}^{\mu \lbrack
t]}a_{R^{\prime }}^{\mu \lbrack t]}\right\rangle _{t}\left\langle
n_{R}^{\underline{S}[t]}n_{R^{\prime }}^{
\underline{S}[t]}\right\rangle _{t} \\ &=&\sum_{R=1}^{2P}\left\langle \left(
n_{R}^{\underline{S}[t]}\right)
^{2}-n_{R}^{\underline{S}[t]}n_{\overline{R}}^{\underline{S}[t]}\right\rangle
_{t}+\sum_{R\neq R^{\prime }\neq \overline{R}}^{2P}\left( 
\frac{1}{P}\sum_{\mu =0}^{P-1}a_{R}^{\mu \lbrack t]}a_{R^{\prime }}^{\mu
\lbrack t]}\right) \left\langle n_{R}^{\underline{S}[t]}n_{R^{\prime
}}^{\underline{S}[t]}\right\rangle _{t}  \notag \\ &=&\sum_{R=1}^{2P}\left\langle
\left( n_{R}^{\underline{S}[t]}\right)
^{2}-n_{R}^{\underline{S}[t]}n_{\overline{R}}^{\underline{S}[t]}\right\rangle
_{t}.  \notag
\end{eqnarray}
This sum over $2P$ terms can however be written as a sum over $P$
terms,
\begin{eqnarray}
\sigma _{\Omega }^{2} &=&\sum_{R=1}^{2P}\left\langle \left(
n_{R}^{\underline{S}[t]}\right)
^{2}-n_{R}^{\underline{S}[t]}n_{\overline{R}}^{\underline{S}[t]}\right\rangle _{t} 
\label{requ} \\ &=&\sum_{R=1}^{P}\left\langle
\left( n_{R}^{\underline{S}[t]}\right)
^{2}-n_{R}^{\underline{S}[t]}n_{\overline{R}}^{\underline{S}[t]}+\left(
n_{\overline{R}}^{\underline{S}[t]}\right)
^{2}-n_{\overline{R}}^{\underline{S}[t]}n_{R}^{\underline{S}[t]}\right\rangle _{t} 
\notag \\ &=&\sum_{R=1}^{P}\left\langle \left( n_{R}^{\underline{S}[t]}\right)
^{2}-2n_{R}^{\underline{S}[t]}n_{\overline{R}}^{\underline{S}[t]}+\left(
n_{\overline{R}}^{\underline{S}[t]}\right) ^{2}\right\rangle _{t}  \notag \\
&=&\sum_{R=1}^{P}\left\langle \left(
n_{R}^{\underline{S}[t]}-n_{\overline{R}}^{\underline{S}[t]}\right)
^{2}\right\rangle _{t} \equiv \left\langle \sum_{R=1}^{P}\left(
n_{R}^{\underline{S}[t]}-n_{\overline{R}}^{\underline{S}[t]}\right)
^{2}\right\rangle _{t}.  \notag
\end{eqnarray}
All the above is for a given realization of the quenched disorder
$\Omega $. The values of $n_{R}^{\underline{S}[t]}$ and
$n_{\overline{R}}^{\underline{S}[t]}$\ for each $R$ will depend on the precise form
of $\Omega $. 

We will now proceed to consider the ensemble-average 
over all possible realizations of quenched disorder. 
The ensemble-average is denoted as $\left\langle ...\right\rangle _{\Omega }$, and
for simplicity the notation $\left\langle \sigma _{\Omega }^{2}\right\rangle
_{\Omega }=\sigma ^{2}$ is defined. This ensemble-average is performed on either
side of Equation 7,
\begin{equation}
\sigma ^{2}=\left\langle \left\langle \sum_{R=1}^{P}\left(
n_{R}^{\underline{S}[t]}-n_{\overline{R}}^{\underline{S}[t]}\right)
^{2}\right\rangle _{t}\right\rangle _{\Omega }
\end{equation}yielding the volatility in the attendance $A(t)$.
However as in Figure 2, the numerical simulations are typically performed
for the volatility of the number of traders choosing option $1$ (e.g. buy).
Fortunately a simple relationship can be derived between the two as follows. Going
back to the original definitions given for the attendance (Equation 1) and
substituting into Equation
\ref{vol2}, gives
\begin{equation}
\sigma ^{2}=\left\langle \left\langle \left[ n_{1}(t)-n_{0}(t)\right]
^{2}\right\rangle _{t}\right\rangle _{\Omega }.
\end{equation}
It is known that the total number of traders $N=n_{1}(t)+n_{0}(t)$,
hence
\begin{equation}
\sigma ^{2}=\left\langle \left\langle \left[ 2n_{1}(t)-N\right] ^{2}\right\rangle
_{t}\right\rangle _{\Omega }.
\end{equation}
By the symmetry of the game however, it is expected that
\begin{equation}
\left\langle \left\langle \left[ n_{1}(t)\right] \right\rangle _{t}\right\rangle
_{\Omega }=\left\langle \left\langle \left[ n_{0}(t)\right]
\right\rangle _{t}\right\rangle _{\Omega }=\frac{N}{2}
\end{equation}
and hence
\begin{eqnarray}
\sigma ^{2} &=&\left\langle \left\langle \left[ 2n_{1}(t)-2\left\langle
\left\langle \left[ n_{1}(t)\right] ^{2}\right\rangle _{t}\right\rangle _{\Omega
}\right] ^{2}\right\rangle _{t}\right\rangle _{\Omega } \\ &=&4\left\langle
\left\langle \left[ n_{1}(t)-\left\langle \left\langle
\left[ n_{1}(t)\right] ^{2}\right\rangle _{t}\right\rangle _{\Omega }\right]
^{2}\right\rangle _{t}\right\rangle _{\Omega }  \notag \\ &=&4\sigma _{1}^{2}. 
\notag
\end{eqnarray}
This leaves the result that the ensemble and time-averaged
volatility of the number of agents choosing a given option is given
by $\sigma _{1}$, where 
\begin{equation}
\sigma _{1}^{2}=\frac{1}{4}\left\langle \left\langle \sum_{R=1}^{P}\left(
n_{R}^{\underline{S}[t]}-n_{\overline{R}}^{\underline{S}[t]}\right)
^{2}\right\rangle _{t}\right\rangle _{\Omega }.  \label{ModelC1}
\end{equation} Equation \ref{ModelC1} is an important intermediary result for the
Crowd-Anticrowd theory. Before proceeding to treat it analytically, it is important
to evaluate it numerically to see how well it describes the numerical results of the
Savit curve. If this is successful, we will have provided a
proof-of-concept of the Crowd-Anticrowd approach. In particular, it will give us
confidence that the Crowd-Anticrowd approach hasn't thrown out any of the essential
physics so far, and reassure us that it is worth proceeding with an analytic
evaluation of Equation 13. 

Figure 7 confirms that the Crowd-Anticrowd theory of Equation 13 does indeed work.
The solid curves represent the ensemble and time-averaged standard deviation of the
number of agents choosing a given option, obtained by recording the number of agents
choosing option
$1$ at every timestep.
The dashed curves show the ensemble and time-averaged standard deviation calculated
using Equation 13. (For the purpose of calculation, the numbers of agents using each
$R$'th ranked strategy were recorded at each timestep for $1000$ turns of the game
after initial transient effects had died down. The volatility for a run of the game
was hence calculated over this time period. Finally an average was taken over $16$
runs of the game to simulate the configuration averaging in Equation 13). As can be
seen from Figure 7, the results show that Equation 13 captures the essential physics
underlying the fluctuations in the Minority Game.

\subsection{Quantitative theory}

Equation 13 provides us with
an exact theory for the time-averaged fluctuations in the MG. However
some form of approximation must be introduced in order to reduce Equation 13 to an
analytic expression. It turns out that Equation 13 can be manipulated in a variety of
ways, depending on the level of approximation that one is prepared to make. The
precise form of any resulting analytic expression will obviously depend on the
details of the approximations made. 

In this Section, we will approach the problem of evaluating Equation 13
analytically by first relabelling strategies. Specifically, the sum in Equation
\ref{ModelC1} is re-written to be over a \emph{virtual-point ranking} $K$\ and not
the decimal form
$R$. Consider the variation in points for a given strategy, as a function of time for
a given realization of the quenched disorder $\Omega $. Figure 8 provides a schematic
representation of how the scores of three such strategies, and their three
anti-correlated strategies, might vary in time (particularly for lower
$m$). The ranking (i.e. label) of a given
strategy in terms of virtual-points score is changing all the time since the
individual strategies have a variation in virtual-points which varies rapidly (see
e.g. the black curve in Figure 8). This implies that the specific identity of the
`n'th highest-scoring strategy' is changing all the time. It also implies that
$n_{R}^{\underline{S}[t]}$ is changing rapidly in time. In order to proceed, we
shift the focus onto the time-evolution of the highest-scoring strategy, second
highest-scoring strategy etc. This has a much smoother time-evolution than the
time-evolution for a given strategy
$S_{R}[t]$. In short, \emph{the focus is shifted from the time-evolution of the
points of a given strategy (i.e. from} $S_{R}[t]$ \emph{) to the time-evolution of
the points of the n'th highest scoring strategy (i.e. to }$S_{K}[t]$\emph{)}.
From this point of view, Figure 8 should now be viewed in terms of virtual-point
ranking $K$. Figure 9 is a schematic representation of how the scores of the two top
scoring strategies from Figure 8 vary, using the new virtual-point ranking scheme. 
The label $K$ is used to denote the rank in terms of strategy score, i.e. $K=1$ is
the highest scoring strategy position, $K=2$ is the second highest-scoring strategy
position etc. with
\begin{equation} S_{K=1}>S_{K=2}>S_{K=3}>S_{K=4}>...
\end{equation}
A given strategy, e.g. $0000$, may at a given timestep have label
$K=1$, while a few timesteps later have label $K=5$. Because it is known that
$S_{R}=-S_{\overline{R}}$\ (i.e. strategy scores start off all at zero), then we
know that
$S_{K}=-S_{\overline{K}}$. Equation 13 can hence be rewritten exactly
as
\begin{equation}
\sigma _{1}^{2}=\frac{1}{4}\left\langle \left\langle \sum_{K=1}^{P}\left(
n_{K}^{\underline{S}[t]}-n_{\overline{K}}^{\underline{S}[t]}\right)
^{2}\right\rangle _{t}\right\rangle _{\Omega }.  \label{ModelC1K}
\end{equation} Note that the quantities $n_{K}^{\underline{S}[t]}$\ and
$n_{\overline{K}}^{\underline{S}[t]}$\ will fluctuate in time, but far less so than
the individual strategy quantities $n_{R}^{\underline{S}[t]}$\ and
$n_{\overline{R}}^{\underline{S}[t]}$ in Equation \ref{ModelC1}. As can be seen from
Figure 9 (e.g. black curve), the time-evolution of the strategy scores is such that
the points for a given $K$ tend to fluctuate around a mean value. It is now assumed
that the spread in traders across the strategy space is fairly uniform, i.e. $\Omega
$ is a fairly uniform matrix. This will be reasonable for small $m$. Hence it is
expected that the number of traders playing the strategy in position $K$ at any
timestep $t$, will also fluctuate around some mean value:
\begin{equation} n_{K}^{\underline{S}[t]}=n_{K}+\varepsilon _{K}(t),
\end{equation}
where $\varepsilon _{K}(t)$\ is assumed to be a white noise term
with zero mean and small variance. Here $n_{K}$\ is the mean value. Hence,
\begin{eqnarray}
\sigma _{1}^{2} &=&\frac{1}{4}\left\langle \sum_{K=1}^{P}\left\langle \left[
n_{K}+\varepsilon _{K}(t)-n_{\overline{K}}-\varepsilon _{\overline{K}}(t)
\right] ^{2}\right\rangle _{t}\right\rangle _{\Omega } \\ &=&\frac{1}{4}\left\langle
\sum_{K=1}^{P}\left\langle \left[ (n_{K}-n_{\overline{K}})+(\varepsilon
_{K}(t)-\varepsilon _{\overline{K}}(t))\right] ^{2}\right\rangle _{t}\right\rangle
_{\Omega }  \notag \\ &=&\frac{1}{4}\left\langle
\sum_{K=1}^{P}\left\langle \left[ n_{K}-n_{\overline{K}}\right] ^{2}+\left[
\varepsilon _{K}(t)-\varepsilon _{\overline{K}}(t)\right] ^{2}+\left[
2(n_{K}-n_{\overline{K}})(\varepsilon _{K}(t)-\varepsilon _{\overline{K}}(t))\right]
\right\rangle _{t}\right\rangle _{\Omega }  \notag \\ &\thickapprox
&\frac{1}{4}\left\langle \sum_{K=1}^{P}\left\langle \left[
n_{K}-n_{\overline{K}}\right] ^{2}\right\rangle _{t}\right\rangle _{\Omega }=
\frac{1}{4}\left\langle \sum_{K=1}^{P}\left[ n_{K}-n_{\overline{K}}\right]
^{2}\right\rangle _{\Omega },  \notag
\end{eqnarray}
since the latter two terms involving noise will average out to be
small. The resulting expression involves no time dependence. The averaging over
$\Omega$ can then be taken inside the sum. The individual terms in the sum, i.e.
$\left\langle \left[ n_{K}-n_{\overline{K}}\right] ^{2}\right\rangle _{\Omega }$,
are just an expectation value of a function of two variables $n_{K}$ and
$n_{\overline{K}}$. Each term can therefore be rewritten exactly using the joint
probability distribution for $n_{K}$ and $n_{\overline{K}}$, which we shall call
$P(n_{K},n_{\overline{K}})$. Hence,
\begin{eqnarray}
\sigma _{1}^{2} &=&\frac{1}{4}\sum_{K=1}^{P}\left\langle \left[
n_{K}-n_{\overline{K}}\right] ^{2}\right\rangle _{\Omega }  \label{general} \\
&=&\frac{1}{4}\sum_{K=1}^{P}\sum_{n_{K}=0}^{N}\sum_{n_{\overline{K}}=0}^{N}
\left[ n_{K}-n_{\overline{K}}\right] ^{2}P(n_{K},n_{\overline{K}}),  \notag
\end{eqnarray}
where the standard probability result involving functions of two
variables has been used. So how can we evaluate Equation 18? In general, it will
depend on the detailed form of the joint probability function
$P(n_{K},n_{\overline{K}})$ which in turn will depend on the ensemble of quenched
disorders $\{\Omega \}$ which are being averaged over. 

We will start off by looking at Equation 18 in the limiting case where the averaging
over the quenched disorder matrix is dominated by the matrices $\Omega $ which are
nearly flat. This will be a good approximation for small $m$ since in this limit the
standard deviation of an element in $\Omega$ (i.e. the standard deviation in
bin-size) is much smaller than the mean bin-size. In this limiting case, there are
several nice features:

\bigskip

\begin{itemize}
\item in addition to the ranking in terms of virtual-points, i.e.
$S_{K=1}>S_{K=2}>S_{K=2}>S_{K=2}>...$ (this holds by definition of the labels 
$\{K\}$), we will also have
\begin{equation} n_{K=1}>n_{K=2}>n_{K=2}>n_{K=2}>...  \notag
\end{equation} [Note that the ordering in terms of the labels $\{R\}$ would not be
sequential, i.e. it is \emph{not} true that $n_{R=1}>n_{R=2}>n_{R=2}>n_{R=2}>...$]
Hence the rankings in terms of highest virtual-points and popularity are identical.

\item it is guaranteed that the strategy $\overline{K}$, which is anticorrelated to
strategy $K$, occupies position $\overline{K}=2P+1-K$\ in this popularity-ranked
list.

\item the probability distribution $P(n_{K},n_{\overline{K}})$ will be sharply
peaked around the $n_{K}$ and $n_{\overline{K}}$ values given by the expected values
for a flat quenched-disorder matrix $\Omega $. We will call these values
$\overline{n_{K}}$ and $\overline{n_{\overline{K}}}$.
\end{itemize}

\bigskip

\noindent The last point implies that $P(n_{K},n_{\overline{K}})\varpropto \delta
(n_{K}-\overline{n_{K}})\delta (n_{\overline{K}}-\overline{n_{\overline{K}}}) $
and so
\begin{equation}
\sigma _{1}^{2}=\frac{1}{4}\sum_{K=1}^{P}\left[
\overline{n_{K}}-\overline{n_{\overline{K}}}\right] ^{2}.  \label{final}
\end{equation} 
We note that there is a very simple interpretation of  Equation
\protect\ref{final}. It represents the sum of the variances for each crowd-anticrowd
pair. For a given strategy $K$ there is an anticorrelated strategy $\overline{K}$.
The $ n_{K}$\ agents using strategy $K$\ are doing the
\emph{opposite} to the $n_{
\overline{K}}$\ agents using strategy $\overline{K}$\ \emph{irrespective} of the
history bit-string. Hence the effective group-size for each crowd-anticrowd pair is
$n_K^{eff}=\overline{n_{K}}-\overline{n_{\overline{K}}}
$: this represents the net step-size $d$ of the crowd-anticrowd pair in a random-walk
contribution to the total variance. Hence, the net contribution by this
crowd-anticrowd pair to the variance is given by
\begin{eqnarray}
\lbrack \sigma _{1}^{2}]_{K\overline{K}} &=&\frac{1}{4}[\sigma ^{2}]_{K
\overline{K}}=\frac{1}{4}.4pqd^{2}=pqd^{2} \\
&=&\frac{[n_K^{eff}]^{2}}{4}=\frac{1}{4}\left[ \overline{n_{K}}-\overline{n_{
\overline{K}}}\right] ^{2}.  \notag
\end{eqnarray}
where $p=q=1/2$ for a random walk. Since all the strong correlations have been
removed (i.e. anti-correlations) it can be happily assumed that the separate
crowd-anticrowd pairs execute random walks which are \emph{uncorrelated} with
respect to each other. [Recall the properties of the RSS - all the remaining
strategies are uncorrelated]. Hence the total variance is given by the sum of the
individual variances,
\begin{equation}
\sigma _{1}^{2}=\sum_{K=1}^{P}[\sigma _{1}^{2}]_{K\overline{K}}=\frac{1}{4}
\sum_{K=1}^{P}\left[ \overline{n_{K}}-\overline{n_{\overline{K}}}\right] ^{2},
\end{equation}
which corresponds exactly to Equation \ref{final}.

\section{Limiting cases of Crowd-Anticrowd \\ theory}

\subsection{Flat quenched disorder matrix $\Omega $, low $m$}

Explicit expressions for the case of a flat quenched disorder matrix $\Omega 
$ can now be calculated. In this limit each element of $\Omega $ has a mean of
$N/(2P)^{s}$ agents per `bin'. For the case $s=2$, the expected number of traders
whose highest scoring strategy is the strategy occupying position $K$ at timestep
$t$, will therefore be given by summing the appropriate rows and columns of this
quenched disorder matrix $\Omega $. The matrix $\Omega $ is flat, so any re-ordering
has no effect on the form of the matrix. Figure 10 provides a schematic
representation of $\Omega $ with $m=2$, $s=2$, in the RSS. These strategies are now
ranked according to the value of $K$. The shaded elements represent those agents
which hold a strategy that is ranked 4'th highest in score, i.e. $K=4$. Any trader
using the strategy in position $K=4$ cannot have any strategy with a higher
position, by definition of the rules of the game. (The traders use their highest
scoring strategy). Hence the traders using the strategy in position $K=4$ must lie
in one of the shaded bins. Since it is assumed that the coverage of the bins is
uniform, the expected number of agents using the strategy in position $K=4$ is given
by
\begin{eqnarray}
\overline{n_{K=4}} &=&N.\frac{1}{(2P)^{2}}\sum (shaded\text{ }bins) \\
&=&N.\frac{1}{64}.[(8-3)+(8-3)-1]  \notag \\ &=&\frac{9}{64}N.  \notag
\end{eqnarray}
For more general $m$ and $s$ values this becomes
\begin{eqnarray}
\overline{n_{K}} &=&\frac{N}{(2P)^{s}}[s(2P-K)^{s-1}+\frac{s(s-1)}{2}
(2P-K)^{s-2}+...+1]  \label{YofR} \\
&=&\frac{N}{(2P)^{s}}\sum_{r=0}^{s-1}\frac{s!}{(s-r)!r!}[2P-K]^{r}  \notag \\
&=&\frac{N}{(2P)^{s}}([2P-K+1]^{s}-[2P-K]^{s})  \notag \\ &=&N.\left( \left[
1-\frac{(K-1)}{2P}\right] ^{s}-\left[ 1-\frac{K}{2P}
\right] ^{s}\right) ,  \notag
\end{eqnarray}
with $P\equiv 2^{m}$. In the case where each agent
holds two strategies, $s=2,$ $\overline{n_{K}}$ can be simplified to
\begin{eqnarray}
\overline{n_{K}} &=&N.\left( \left[ 1-\frac{(K-1)}{2P}\right] ^{2}-\left[ 1-
\frac{K}{2P}\right] ^{2}\right)  \label{k} \\
&=&\frac{(2^{m+2}-2K+1)}{2^{2(m+1)}}N.  \notag
\end{eqnarray}
Similarly for $\overline{n_{\overline{K}}}$ the simplification is as
follows:
\begin{eqnarray}
\overline{n_{\overline{K}}} &=&\frac{(2^{m+2}-2\overline{K}+1)}{2^{2(m+1)}}N
\label{kbar} \\ &=&\frac{(2K-1)}{2^{2(m+1)}}N,  \notag
\end{eqnarray}
where the relation $\overline{K}=2P-K+1\equiv 2^{m+1}-K+1$\ is used.
It is emphasized that these results depend on the assumption that the averages are
dominated by the effects of flat distributions for the quenched disorder matrix
$\Omega $, and hence will only be quantitatively valid for low $m$.

Using Equations \ref{k}\ and \ref{kbar}\ in Equation \ref{final} gives
\begin{eqnarray}
\sigma _{1}^{2} &=&\frac{1}{4}\sum_{K=1}^{P}\left[ \frac{(2^{m+2}-2K+1)}{
2^{2(m+1)}}N-\frac{(2K-1)}{2^{2(m+1)}}N\right] ^{2} \\
&=&\frac{N^{2}}{2^{4(m+1)}}\sum_{K=1}^{P}[2^{m+1}-2K+1]^{2}  \notag \\
&=&\frac{N^{2}}{3.2^{m+2}}(1-2^{-2(m+1)}),  \notag
\end{eqnarray}
and hence
\begin{equation}
\sigma _{1}^{delta\text{ }f}=\frac{N}{\sqrt{3}.2^{\frac{m}{2}+1}}
(1-2^{-2(m+1)})^{\frac{1}{2}},  \label{flatom}
\end{equation} which is valid for small $m$. Numerical results show that this is
indeed the case. (The rationale behind the choice of superscript will become apparent
shortly.)

\subsection{Non-flat quenched disorder $\Omega $ at low $m$}

The appearance of a significant number of non-flat quenched disorder matrices
$\Omega $ in the ensemble, implies that the standard deviation for each `bin' is now
significant, i.e. non-negligible compared to the mean. This will be increasingly
true as $m$ increases. In this case, the general analysis is much more complicated,
and should really appeal to the dynamics. However, an approximate theory which gives
good agreement with the numerical results can be developed. The features for the
case of ensembles containing a significant number of non-flat quenched disorder
matrices $\Omega $ are as follows:

\begin{itemize}
\item  By definition of the labels $\{K\}$, the ranking in terms of virtual-points
is retained, i.e. S$_{K=1}>$S$_{K=2}>$S$_{K=3}>$S$_{K=4}>...$ is always true.
However, the disorder in the matrix $\Omega $ distorts the number of agents playing
a given strategy away from the flat-matrix results. Hence it is not in general true
that $n_{K=1}>n_{K=2}>n_{K=3}>n_{K=4}>...$, and hence the rankings in terms of
highest virtual-points and popularity are no longer identical.

\item Instead we have that $n_{K^{^{\prime }}}>n_{K^{^{\prime \prime
}}}>n_{K^{^{\prime
\prime \prime }}}>n_{K^{^{\prime \prime \prime \prime }}}>...$, where the label
$K^{\prime }$ need not equal $1$, and $K^{\prime \prime }$ need not equal $2$ etc..
It is however possible to introduce a new label $\{Q\}$ which will rank the
strategies in terms of popularity, i.e.
\begin{equation*} n_{Q=1}>n_{Q=2}>n_{Q=3}>n_{Q=4}>...,
\end{equation*} where $Q=1$ represents $K^{\prime }$, $Q=2$ represents $K^{\prime
\prime }$, etc.
\end{itemize} 

\noindent With this in mind, we will return to the original general form for the
volatility  in Equation \ref{general}, but rewrite it slightly as follows:
\begin{eqnarray}
\sigma _{1}^{2} &=&\frac{1}{4}\sum_{K=1}^{P}\sum_{n_{K}=0}^{N}\sum_{n_{
\overline{K}}=0}^{N}\left[ n_{K}-n_{\overline{K}}\right] ^{2}P(n_{K},n_{
\overline{K}}) \\
&=&\frac{1}{8}\sum_{K=1}^{2P}\sum_{n_{K}=0}^{N}\sum_{n_{\overline{K}}=0}^{N}
\left[ n_{K}-n_{\overline{K}}\right] ^{2}P(n_{K},n_{\overline{K}})  \notag \\
&=&\frac{1}{8}\sum_{K=1}^{2P}\sum_{K^{\prime }=1}^{2P}\left\{
\sum_{n_{K}=0}^{N}\sum_{n_{K^{\prime }}=0}^{N}\left[ n_{K}-n_{K^{\prime }}
\right] ^{2}P(n_{K},n_{K^{\prime }})\right\} f_{K^{\prime },\overline{K}}, 
\notag
\end{eqnarray}
where the probability that $K^{\prime }=\overline{K}$\ is given by $
f_{K^{\prime },\overline{K}}=\delta _{K^{\prime }=\overline{K}}$ and $
\overline{K}=2P+1-K$. So far, this manipulation is exact.

A switch is now made to the popularity-labels $\{Q\}$. Consider any particular
strategy which was labelled previously by $K$ and is now labelled by $Q$. Unlike the
case of the flat disorder matrix, it is \emph{not} guaranteed that this strategy's
anticorrelated partner will lie in position $\overline{Q}=2P+1-Q$. This is because
of the relabelling operation: all that can be said is that the strategy $R$ has
changed label from $K\rightarrow Q(K)$ while the anticorrelated strategy has changed
label from $\overline{K}
\rightarrow \overline{Q}(\overline{K})$ and that in general $\overline{Q}
\neq 2P+1-Q$.

As a result of the relabelling, $n_{Q=1}>n_{Q=2}>n_{Q=3}>n_{Q=4}>...$ as stated
earlier. Assuming that the main effect of the non-flat matrix was to shuffle these
numbers, as opposed to altering their values, it can be assumed that as a
zeroth-order approximation the values of $n_{Q=1}$, $n_{Q=2}$,
$n_{Q=3,\text{...}}$ etc. are still sharply peaked around the expected values
obtained for the flat-matrix case, i.e. it is assumed that the probability
distribution $P(n_{Q(K)},n_{Q^{\prime }(\overline{K})})$ will be sharply peaked
around the $n_{Q(K)}$\ and $n_{Q^{\prime }(\overline{K })}$ values given by the
expected values for the (nearly flat) quenched-disorder matrix $\Omega $. Lets call
these values $\overline{n_{Q}}$ and $\overline{n_{Q^{\prime }}}$ where the intrinsic
dependence of $Q$ on $K$ has been dropped. Hence $P(n_{Q},n_{Q^{\prime }})\varpropto
\delta (n_{Q}-
\overline{n_{Q}})\delta (n_{Q^{\prime }}-\overline{n_{Q^{\prime }}})$ with $
\overline{n_{Q}}$ and $\overline{n_{Q^{\prime }}}$ given by the box-counting method
(see Figure 10 but with $K,K'$ replaced by $Q,Q'$). The volatility becomes, after
relabelling:
\begin{equation}
\sigma _{1}^{2}=\frac{1}{8}\sum_{Q=1}^{2P}\sum_{Q^{\prime }=1}^{2P}\left\{
\sum_{n_{Q}=0}^{N}\sum_{n_{Q^{\prime }}=0}^{N}\left[ n_{Q}-n_{Q^{\prime }}
\right] ^{2}P(n_{Q},n_{Q^{\prime }})\right\} f_{Q^{\prime },\overline{Q}},
\end{equation}
where $f_{Q^{\prime },\overline{Q}}$\ is the probability that
strategy with label $Q^{\prime }$ is anticorrelated to $Q$. Substituting in $
P(n_{Q},n_{Q^{\prime }})\varpropto \delta (n_{Q}-\overline{n_{Q}})\delta
(n_{Q^{\prime }}-\overline{n_{Q^{\prime }}})$ gives
\begin{equation}
\sigma _{1}^{2}=\frac{1}{8}\sum_{Q=1}^{2P}\sum_{Q^{\prime }=1}^{2P}\left[ 
\overline{n_{Q}}-\overline{n_{Q^{\prime }}}\right] ^{2}f_{Q^{\prime },
\overline{Q}},  \label{fq}
\end{equation}
where the function $f_{Q^{\prime },\overline{Q}}$, which is the
probability that the strategy with label $Q^{\prime }$ is the anticorrelated
strategy $
\overline{Q}$, still needs to be specified.

So what should the form of $f_{Q^{\prime },\overline{Q}}$ be? In principle, it
should include the effects of market impact/dynamical feedback which develops as the
game progresses, in addition to the effects of the disorder. An exact expression for
this term is yet to be found. However, two possibilities have been tried:
\\

\noindent (i) assume a form that depends on $m$ and $N$.  This approach was first
pursued in Ref. \cite{CAT1} and gives good analytic agreement with the numerical
simulations (see later Figure 15). It will be discussed shortly.
\\

\noindent (ii) assume that the probability that $Q^{\prime }$ is the anticorrelated
strategy $\overline{Q}$, is given by $1/(2P)$ and is \emph{independent} of the label
$Q^{\prime }$. In this sense this is the 
opposite limit to the delta-function case for flat disorder
matrix. Instead of being a delta-function at $Q^{\prime }=\overline{Q}=2P+1-Q$, it
is said that the anticorrelated strategy $\overline{Q}$ could be anywhere in the
list of $1...2P$ strategies.
\\

\noindent Case (ii) is the first approach we will pursue here. In this limiting case
of
$1/(2P)$, one obtains
\begin{equation}
\sigma _{1}^{flat\text{
}f}=\frac{N}{\sqrt{3}.2^{(m+3)/2}}(1-2^{-2(m+1)})^{\frac{1}{2}}.  \label{nonflatom}
\end{equation} Comparing this with Equation \ref{flatom} it can be seen that
\begin{equation}
\sigma _{1}^{flat\text{ }f}=\frac{1}{\sqrt{2}}\sigma _{1}^{delta\text{ } f}\approx
0.7\sigma _{1}^{delta\text{ }f}.
\end{equation} for the case of $s=2$. It should be noted that Equation \ref{flatom}
can be derived from Equation \ref{fq} by letting $f_{Q^{\prime },\overline{Q}}$\
take a $\delta $-function distribution $\delta _{Q^{\prime },2^{m+1}+1-Q}$ peaked at
$Q^{\prime }=2^{m+1}+1-Q$, hence the superscript in Equation 27.

\subsection{Non-flat quenched disorder $\Omega $ at high $m$}

For the limit of high $m$, the granular nature of the disorder becomes important. In
other words, the standard deviation in the number of traders in a given bin is now
similar to the mean value hence the fluctuations (which are limited to integer
numbers) are important. Note that for large $m$, these integer numbers tend to be
$0$'s and $1$'s for each box $(Q,Q^{\prime })$. It almost looks like the problem
of fermions in energy levels - double occupancy does not occur. In this limit of
high $m$ (by high $m$ it is meant that the number of strategies is greater than
$N.s$, i.e. $2.2^{m}>N.s$) there will be $N$ crowds each representing one strategy
and one agent. In the limit that $f_{Q^{\prime },\overline{Q}}=1/(2P)$, the
probability that a strategy $Q$ representing one agent is matched with its
anticorrelated strategy in position $\overline{Q}$ also representing one agent, is
given by $N/2P$. The probability that a strategy $Q$ representing one agent is
matched with its anticorrelated strategy in position $\overline{Q}$ which represents
a strategy that no agent is using, is given by $1-N/2P$. Using Equation \ref{fq}
then gives,
\begin{eqnarray}
\sigma _{1}^{2} &=&\frac{1}{8}\sum_{Q=1}^{2P}\sum_{Q^{\prime }=1}^{2P}\left[ 
\overline{n_{Q}}-\overline{n_{Q^{\prime }}}\right] ^{2}f_{Q^{\prime },
\overline{Q}} \\ &=&\frac{1}{4}\sum_{Q=1}^{N}\left\{ \left[
(\overline{n_{Q}}=1)-(\overline{n_{Q^{\prime }}}=1)\right] ^{2}\frac{N}{2P}+\left[
(\overline{n_{Q}}=1)-(
\overline{n_{Q^{\prime }}}=0)\right] ^{2}\frac{2P-N}{2P}\right\} ,  \notag
\end{eqnarray}
where the sum is now performed over the $N$ strategies that each
have one agent subscribed, and the second summation over $Q^{\prime }$ has been
absorbed by the extra probabilistic factors. Simplifying this expression gives,
\begin{eqnarray}
\sigma _{1}^{flat\text{ }f,\text{ }high\text{ }m} &=&\left( \frac{1}{4}
\sum_{Q=1}^{N}\left[ (\overline{n_{Q}}=1)-(\overline{n_{Q^{\prime }}}=0)
\right] ^{2}\frac{2P-N}{2P}\right) ^{\frac{1}{2}}  \label{highm} \\
&=&(\frac{1}{4}N.\frac{2P-N}{2P})^{\frac{1}{2}}  \notag \\
&=&\frac{\sqrt{N}}{2}(1-\frac{N}{2^{m+1}})^{\frac{1}{2}},  \notag
\end{eqnarray}
where $P\equiv 2^{m}$ has been used. It should be noted that $\sigma
_{1}^{delta\text{ }f}$\ , $\sigma _{1}^{flat
\text{ }f}$\ and$\ \sigma _{1}^{flat\text{ }f,\text{ }high\text{ }m}$, are only
limits with regards to the way that the agents are distributed amongst the elements
in the quenched disorder matrix $\Omega $. They do not provide strict bounds
on the actual standard deviation of the numbers of agents choosing a given option
in the simulation.

Figure 2 shows (small circles) $\sigma _{1}$\ as measured from the simulation
for individual quenched disorder matrices $\Omega $, as a function of agent memory
size
$m$. The spread in values from individual runs, for a given $m$, indicates the
extent to which the choice of $\Omega
$\ alters the dynamics of the MG. The upper line at low $m$, is Equation
\ref{flatom}\ showing\
$\sigma _{1}^{delta\text{ }f}$. The lower line at low $m$, is Equation
\ref{nonflatom}\ showing\ $\sigma _{1}^{flat\text{ }f}$. The line at high $m$ is
Equation \ref{highm}
\ showing $\sigma _{1}^{flat\text{ }f,\text{ }high\text{ }m}$. Figure 11 shows
$\sigma _{1}^{delta\text{ }f}$\ , $\sigma _{1}^{flat\text{  }f}$\ and$\ \sigma
_{1}^{flat\text{ }f,\text{ }high\text{ }m}$, as a function of $m$ for $s=2$, $4$ and
$8$. Comparing the analytic curves with the numerical results seen in Figure 11,
it can be seen that these analytic expressions capture the essential physics driving
the variation in behaviour of the volatility. They have been obtained within a static
framework, confirming that the volatility does not depend in a sensitive way on the
details of the dynamics.

\subsection{An investigation of strategy rankings}

We now investigate more carefully the form of $f_{Q^{\prime
},\overline{Q}}$. Recall Equation
\ref{fq},
\begin{equation}
\sigma _{1}^{2}=\frac{1}{8}\sum_{Q=1}^{2P}\sum_{Q^{\prime }=1}^{2P}\left[ 
\overline{n_{Q}}-\overline{n_{Q^{\prime }}}\right] ^{2}f_{Q^{\prime },
\overline{Q}}
\end{equation} The probability distribution $f_{Q^{\prime },\overline{Q}}$ gives the
probability that the $Q^{\prime }$'th most popular strategy is the anticorrelated
strategy to the $Q$'th most popular strategy. It has already been stated that an
exact form for $f_{Q^{\prime },\overline{Q}}$\ has yet to be found.

Figure 12 compares the theoretical values of $n_{Q}$ calculated using Equation
\ref{YofR} for $s=2$ and $N=101$ (dashed lines) to numerical values taken from the
MG simulation (solid lines). We have dropped the bar over $n_{Q}$ for
simplicity, since we are always talking about time-averages. The agreement is good.
In this comparison it is necessary to consider the granular nature of the MG: this
effect becomes increasingly important as the agents' memory size $m$ increases. As
such, the quantities $n_{Q}$ are rounded to the nearest integer to account for the
fact that agents exist only as integer values. This is subject to the constraint that
\begin{equation}
\sum_{Q=1}^{Q=2P}n_{Q}=N.
\end{equation} Only a limited number (call this $B$) terms should be included in
this sum, subject to the condition that the partial sum equals $N$ \emph{after} the
quantities $n_{Q}$ have been rounded to the nearest integer. In addition any
$n_{Q}$'s which are less than one,\ are rounded up to one if $Q\leq B$ such that the
first $B$ terms are all non-zero and sum to $N$. There are hence only $B$
different strategies in play; note that $B\leq 2.2^{m}$ and $B\leq N$. Figure 13
shows $f_{Q^{\prime },\overline{Q}}$ for $Q=1$ as a function of $Q^{\prime }$, taken
from the numerical MG simulation at $m=2,5$ and $10$. We note the following
properties:

\bigskip

\begin{itemize}
\item For small $m$ ($m=2$) the anticorrelated strategy to the most popular strategy
(i.e. $Q=1$) is at $Q^{\prime }=2^{m+1}$, i.e. it is the least popular strategy.
Hence $f_{Q^{\prime },\overline{Q}}$ resembles the $\delta 
$-function limiting case mentioned above. Very few
agents will therefore pick this anticorrelated strategy. Hence the crowd-anticrowd
cancellation will be small and $\sigma _{1}$ will be large, as can be seen in Figures
11 and 2.

\item As $m$ increases ($m=5$) a remarkable effect occurs: the peak in $f_{Q^{\prime
},\overline{Q}}$ moves up toward $Q'=1$. Hence both $Q=1$ and its anticorrelated
partner $\overline{Q}$ are now very popular. Whereas for $m=2$ it seemed like
there was an effective `repulsion' between $Q$ and $\overline{Q}$, for $m=5$ this
now seems more like an attraction. Amusingly, the shape of $f_{Q^{\prime
},\overline{Q}}$ for $m=5$ is now reminiscent of the screening effect of a negative
charge cloud around a positive charge placed at $Q=1$, or even a bound electron-hole
pair (i.e. exciton) with the crowd (anticrowd) playing the role of the positive
(negative) charge. The consequence of this
attraction which appears as $m$ increases, is that the crowd
and anticrowd become comparable in size, yielding a significant cancellation and
hence small volatility as observed in the Savit curve (i.e. Figure 2).

\item For large $m$ ($m=10$), the ability of the anticrowd to `screen' the crowd has
decreased yielding a rather flat distribution as shown. 
\end{itemize}

\bigskip
\noindent Hence $\sigma _{1}$ is small for $m\sim 5-6$, in agreement with Figure 2.
Note that, at intermediate $m$, the MG cannot fully `optimize' itself by building
crowds and anticrowds of exactly equal size. This is due to the initial quenched
disorder
$\Omega $, and the random processes used to resolve the decisions of agents in the
instances when they have strategies with equal past performance. This explains why
the volatility does not go to zero at finite $m$, and why the label of `phase
transition' to separate the small and large $m$ regimes in the Savit curve, is not
strictly correct.

Figure 14 shows the spread of numerical values for different runs of the MG compared
to the crowd-anticrowd theoretical calculation (solid circles) using Equation
\ref{fq}. In contrast to the previous theoretical curves which used analytic
expressions for the probability function
$f_{Q^{\prime },\overline{Q}}$, the
results for each $m$ in Figure 14 have been obtained by generating the
corresponding
$f_{Q^{\prime },\overline{Q}}$ numerically, as in Figure 13. The
agreement is very good. The theoretical points tend to lie toward the high end of
the numerical spread, for example at $m=2$; this can be attributed to the fact that
the theory neglects accidental degeneracies in the virtual-point ordered list
$\{n_{K}\} $. It has been checked that including a stochastic (i.e. coin-toss)
process to break such ties, reduces the theoretical $\sigma $ values down toward the
mid-point of the numerical spread.

\subsection{Using approximate strategy rankings}

We now return to a discussion of case (i) which was referred to earlier, and used in
Ref. \cite{CAT1}. It turns out to be convenient to start with
Equation
\ref{final} for the squared volatility of the number of agents choosing a given
option,
\begin{equation}
\sigma _{1}^{2}=\frac{1}{4}\sum_{K=1}^{P}\left[ \overline{n_{K}}-\overline{
n_{\overline{K}}}\right] ^{2}.
\end{equation}
It has been established that $S_{K=1}>S_{K=2}>S_{K=2}>S_{K=2}>...$
is always true and that for a flat quenched disorder matrix $\Omega $ a similar
statement can be made for the numbers of agents using the $K$'th ranked scoring
strategy, i.e. $n_{K=1}>n_{K=2}>n_{K=2}>n_{K=2}>...$. It is also guaranteed that the
strategy $\overline{K}$, which is anticorrelated to strategy $K$, occupies position
$\overline{K}=2P+1-K$\ in this performance-ranked list. Whilst assuming a flat
quenched disorder matrix $
\Omega $ is valid at low $m$, it is not however valid at high $m$. This is due to
granular effects as the value of $m$ is increased. It remains true that the strategy
$\overline{K}$, which is anticorrelated to strategy $K$, occupies position
$\overline{K}=2P+1-K$\ in the performance-ranked list, however there is no guarantee
that the strategy represented by $\overline{K}$  is even present in the game. It is
possible, and indeed highly likely at high $m$, that the strategy $\overline{K}$ was
not picked by any of the agents at the start of the game. For example if $m=8$,
there are $512$ strategies in the RSS, and for $N=101$ and $s=2$ there will only be
up to $2 \times 101$ of the $512$ possible strategies actually present in the game
(repetition during initial strategy-picking is allowed).

To describe this situation, Ref. \cite{CAT1} introduced a probability
$P(\overline{K}\in
\mathcal{G})$. (We referred to this as case (i) earlier in the present paper). This
corresponds to the following probability: given that strategy
$K$ is used, then
$P(\overline{K}\in
\mathcal{G})$ is the probability that strategy
$\overline{K}$\ is a member of the set of strategies $\mathcal{G}$ that have been
chosen by the agents at the start of the game. Equation 
\ref{final} is rewritten to include this probability: 
\begin{equation}
\sigma _{1}^{2}=\frac{1}{4}\sum_{K=1}^{P}\left[ \overline{n_{K}}-P(\overline{K}\in
\mathcal{G})\overline{n_{\overline{K}}}\right] ^{2}+\frac{1}{4}
\sum_{K=1}^{P}\left[ (1-P(\overline{K}\in \mathcal{G}))\overline{n_{
\overline{K}}}\right] ^{2}.  \label{final2}
\end{equation} The second term in Equation \ref{final2}\ accounts for those
strategies which are not anti-correlated to any other strategy. We now recall
Equation \ref{YofR} which gives the mean number of agents when assuming a flat
quenched disorder matrix $\Omega $: 
\begin{equation}
\overline{n_{K}}=N \left( \left[ 1-\frac{(K-1)}{2P}\right] ^{s}-\left[ 1-
\frac{K}{2P}\right] ^{s}\right)
\end{equation} Taking into account the granular nature of the game, implies that the
condition $\sum_{K=1}^{K=2P}\overline{n_{K}}=N$ should also be imposed. Hence only
$G$ terms in this sum should be included, subject to the condition that the partial
sum equals $N$ \emph{after} the quantities $
\overline{n_{K}}$ have been rounded to the nearest integer. In addition, any 
$\overline{n_{K}}$'s which are less than one, are rounded up to one if $K\leq G$
such that the first $G$ terms are all non-zero. There are hence only $G$ different
strategies in play; note that $G\leq 2\cdot 2^{m}$ and $G\leq N$. The probability
$P(\overline{K}\in \mathcal{G})$ is now the probability that the $[G+1-K]$-th
strategy is anticorrelated to the $K$-th strategy and is a member of the set
$\mathcal{G}$ of strategies chosen by agents at the start of the game. The variance
can hence be written analytically as
\begin{eqnarray}
\sigma _{an}^{2} &=&\frac{1}{4}\sum_{K=1}^{\frac{1}{2}(G-g)}\left[ \overline{
n_{K}}-P(\overline{K}\in \mathcal{G})\overline{n_{G+1-K}}\right] ^{2}
\label{masters} \\ &&+\frac{1}{4}\sum_{K=1}^{\frac{1}{2}(G-g)}\left[
(1-P(\overline{K}\in 
\mathcal{G}))\overline{n_{G+1-K}}\right] ^{2}+\frac{g}{4}[n_{\frac{G+1}{2}}]^{2}, 
\notag
\end{eqnarray}
where $g=0$ if $G$ is even and $g=1$ if $G$ is odd. The first term
represents the net effect after pairing off the agents playing anticorrelated
strategies. The second term in Equation \ref{masters} reintroduces those agents
using strategies that were assumed to be anticorrelated to some more successful
strategy, and hence were discarded unnecessarily in the first term. The third term
in Equation \ref{masters} is due to the volatility of the group which remains
unpaired in the case where the number of different strategies used in the
calculation is odd. The
third term is usually negligible compared to the first two.
$P(\overline{K}\in
\mathcal{G})$ can be approximated by $P(\overline{K}\in \mathcal{G})=p$ for
$%
p<1$ and $P(\overline{K}\in \mathcal{G})=1$ for $p>1$, where
$p=N/(2.2^{m})$%
. Although the form for $P(\overline{K}\in \mathcal{G})$ can be made more
accurate, the present expression is reasonable since there are only of order
$N$ strategies out of a possible maximum of $2.2^{m}$ which can actually be
in play at any one time. Hence, as expected, $P(\overline{K}\in
\mathcal{G})$
is zero when $N<<2.2^{m}$ and unity when $N>>2.2^{m}$. It has been checked
that the analytic results for the volatility are fairly insensitive to the
precise form of $P(\overline{K}\in \mathcal{G})$ as long as $P(\overline{K}%
\in \mathcal{G})$ satisifes the above criteria.

Figure 15 shows the full analytical expression for the volatility curves (solid
lines) of the number of agents choosing a given option for $s=2$, $s=4$ and $s=6$
with $N=101$, compared with the numerical simulations (dashed line). Note that since
$m$ is integer, the curves are not smooth. The
agreement between the analytic and numerical results is good across a wide range of
$m$ and $s$ values. In particular, the analytic results capture the deepening of the
minimum in the volatility as $s$ decreases.

To summarize, these analytic approaches are approximations to
the exact theory of Equation 13. While the approximations adopted have some subtle
differences of emphasis, the resulting analytic expressions do capture
the essential physics underlying the volatility, over a wide range of $m$ and $s$
values: the essence of this physics is the correlations in strategy space and the
associated crowd-anticrowd behavior.

\subsection{Reduced vs. Full Strategy Space}

The volatility of the MG is qualitatively the same when played in both the FSS and
the RSS \cite{RSS}. Quantitatively the volatility of the MG when played using the
RSS is very slightly larger than that of a game played using the FSS. One may
therefore ask why the MG is so similar in characteristics when played in the RSS and
the FSS, and hence why the crowd-anticrowd theory also provides a valid description
for the MG when played in the FSS.

For a game played in the FSS there are $2^{2^{m}}/2.2^{m}$ distinct subsets of
strategies. Each subset can be considered as a separate RSS. Note that the
strategies that belong to a given RSS are optimally spread out across the
corresponding FSS hypercube. Figure 3 shows the distribution of the $16$, $m=2$
strategies across the $4$ dimensional hypercube. The positions of the strategies
belonging to the RSS are such that no two strategies have a Hamming separation less
than $2^{m}/2$. The same can be said for any other choice of RSS. Due to the nature
of a RSS, then given similar strings of past outcomes from which to score strategies
over, each strategy within the RSS attains a score in an uncorrelated or
anticorrelated manner to any other strategy in the subset. Any other RSS within the
FSS will score its strategies in a similar way, although slightly `out of phase'.
For example for $m=3$, the first RSS to be considered could contain the strategy
$00001111$, and the second RSS considered could contain the strategy $01001111 $. It
is easy to see that on $7$ out of $8$ occasions these two strategies would score in
the same way. So, given the nature of the MG (i.e. over a sufficient period of time
in a typical game, any history is just as likely to be followed by a `$0$' or a
`$1$' for low $m$, whilst at high $m$ cooperative effects die off), it can be seen
that these two strategies from two separate RSS follow each other during a typical
run of the game. This argument extends across all strategies in all of the
$2^{2^{m}}/2.2^{m}$ distinct RSS's within the FSS. Hence a game using the FSS
behaves as if there are $2.2^{m}$ clusters of strategies and so is similar to a game
played in the RSS. These clusters form the crowds and anticrowds of the theory and
this clustering allows the use of just one RSS in the analysis of the MG.

We note that the present crowd-anticrowd theory, even within its RSS formulation,
also provides a quantitative theory \cite{generalTherm} to explain the surprising
suppression of volatility observed numerically for the `Thermal Minority Game' (TMG)
\cite{SherThermLett}. In the TMG agents choose between their strategies using an
exponential probability weighting. This reduction in $\sigma $ for stochastic
strategies seems fairly general: for example, the earlier work of Ref. \cite{prl}
provided a modified MG in which agents with stochastic strategies also generate a
smaller-than-random $\sigma$. We have shown
\cite{generalTherm} that incorporating stochastic strategy-use tends to increase
 the crowd-anticrowd cancellation by segregating the population into two opposite
groups, thereby increasing the amount of crowd-anticrowd segregation. This
reduces
$\sigma $ below the random coin-toss limit. In addition we have shown \cite{CAA} how
the theory also works for a mixed population of such `thermal' agents and non-thermal
agents (i.e.
$T=0$). We refer to Refs. \cite{generalTherm} and \cite{CAA} for a detailed
discussion and graphs showing the good quantitative agreement between the
Crowd-Anticrowd theory and the numerical results. We also note that in Ref.
\cite{prl}, which features the MG variant in which agents have purely stochastic
strategies, it is shown that crowd-anticrowd formation arises as a natural consequence
of agents' adaptation in this competitive multi-agent game.

To our knowledge, there is no other analytic theory which can match the range
of quantitative agreement for these models. This suggests that the
Crowd-Anticrowd theory is an important milestone in such multi-agent systems, and
possibly even for Complex Systems in general. It is also pleasing from the point of
view of physics methodology, since the basic underlying philosophy of accounting
correctly for `inter-strategy' correlations is already known to be successful in more
conventional areas of many-body physics. This raises the interesting possibility that
conventional many-body physics might be open to re-interpretation in terms of an
appropriate multi-particle `game': we leave this for future work.

\section{Conclusion}

We have given an in-depth presentation of the Crowd-Anticrowd theory in order to
understand the time-averaged and configuration-averaged fluctuations in the MG
system. The quantitative success of the Crowd-Anticrowd theory means that much of the
time-averaged and ensemble-averaged properties of such MG-like systems can be
understood without having to solve the detailed game dynamics. Since
the theory depends on structure in strategy space, as opposed to the minority
character of the game, we believe that the Crowd-Anticrowd theory will have
applicability for more general multi-agent systems. In particular, we are currently
investigating the modifications that need to be made to the theory when the game is
placed on a network, or there is some form of local connectivity between some (or
all) agents. These findings will be reported elsewhere. 

We believe that the
Crowd-Anticrowd concept might serve as a fundamental theoretical concept for more
general Complex Systems which mimic competitive multi-agent games.
Obviously  there will be some properties of MG games which cannot be described using
such time- and configuration-averaged theories as used here. In particular, any
observation of a real-world Complex System which is thought to resemble a
multi-agent game, will more likely to correspond to a
\emph{single} run which evolves from a specific initial configuration of agents'
strategies. This implies a particular $\Omega$, hence the time-averagings within the
Crowd-Anticrowd theory must be carried out for that particular choice of $\Omega$
\cite{general}. We refer to Refs.
\cite{THMG1,THMG2} for a detailed discussion of such run-specific dynamics. In order
to study time-dependent properties in more general systems, a dynamical form of the
crowd-anticrowd approach can be developed in which the time-dependence of the
crowd-anticrowd sizes is included specifically.

\vskip0.2in
\noindent {\bf Acknowledgements} Much of this work has benefited from our on-going
research collaboration with Paul Jefferies and Pak Ming Hui.

\newpage

{\bf FIGURE CAPTIONS}

\vskip0.1in
\noindent FIGURE 1  Minority Game (MG): at timestep $t$, each agent decides whether
to enter a game where the choices are option $1$ and option $0$. A total of
$n_{0}(t)$\ agents choose $0$ while
$n_{1}(t)$\ choose $1$, with $n_{0}(t)+n_{1}(t)=N$. 

\vskip0.1in
\noindent FIGURE 2  Volatility (small circles) as measured from individual
runs of numerical simulations, as a function of agent memory size $m$ with $s=2$ and
$N=101$. Each run corresponds to a randomly-chosen quenched disorder matrix
$\Omega$. 
Upper line at low
$m$ is Equation \ref{flatom}\ showing\ $\sigma _{1}^{delta\text{ }f}$. Lower line at
low $m$, is Equation \ref{nonflatom}\ showing\ $\sigma _{1}^{flat\text{ }f}$. Line
at high $m$ is Equation
\ref{highm}
\ showing $\sigma _{1}^{flat\text{ }f,\text{ }high\text{ }m}$. Dashed line is random
coin-toss limit.

\vskip0.1in
\noindent FIGURE 3 An $m=2$ strategy space together with some example strategies
(left). The strategy space shown is known as the `Full Strategy Space', FSS, and
contains all possible permutations of the binary options $0$ and $1$ for each
history. There are $2^{2^{m}}$ strategies in this space. The $2^{m}$ dimensional
hypercube (right) shows all 
$2^{2^{m}}$ strategies from the full strategy space at its vertices.

\vskip0.1in
\noindent FIGURE 4 Examples of the de Bruijn graph for $m=1,2,$ and $3$. The
probability that the outcome at time $t+1$ will be a $1$ (or
$0$) depends on the state at time $t$.

\vskip0.1in
\noindent FIGURE 5 Example distribution for the tensor $\Omega$ describing the
strategy allocation for
$N=101$ agents in the case of $m=2$, and $s=2$ in the reduced strategy space RSS.

\vskip0.1in
\noindent FIGURE 6 Qualitative  explanation of the competition between
crowd-anticrowd sizes as a function of $m$. The group of agents whose highest
strategy  is $R$, provides a crowd since they will all use the same strategy at a
particular time-step (and  hence act in the same way regardless of the particular
history bit-string at that timestep). Upper panel: 
run-averaged version of Figure 2, showing the run-averaged volatility of the number
of agents choosing a  particular option as a function of the memory size $m$.

\vskip0.1in

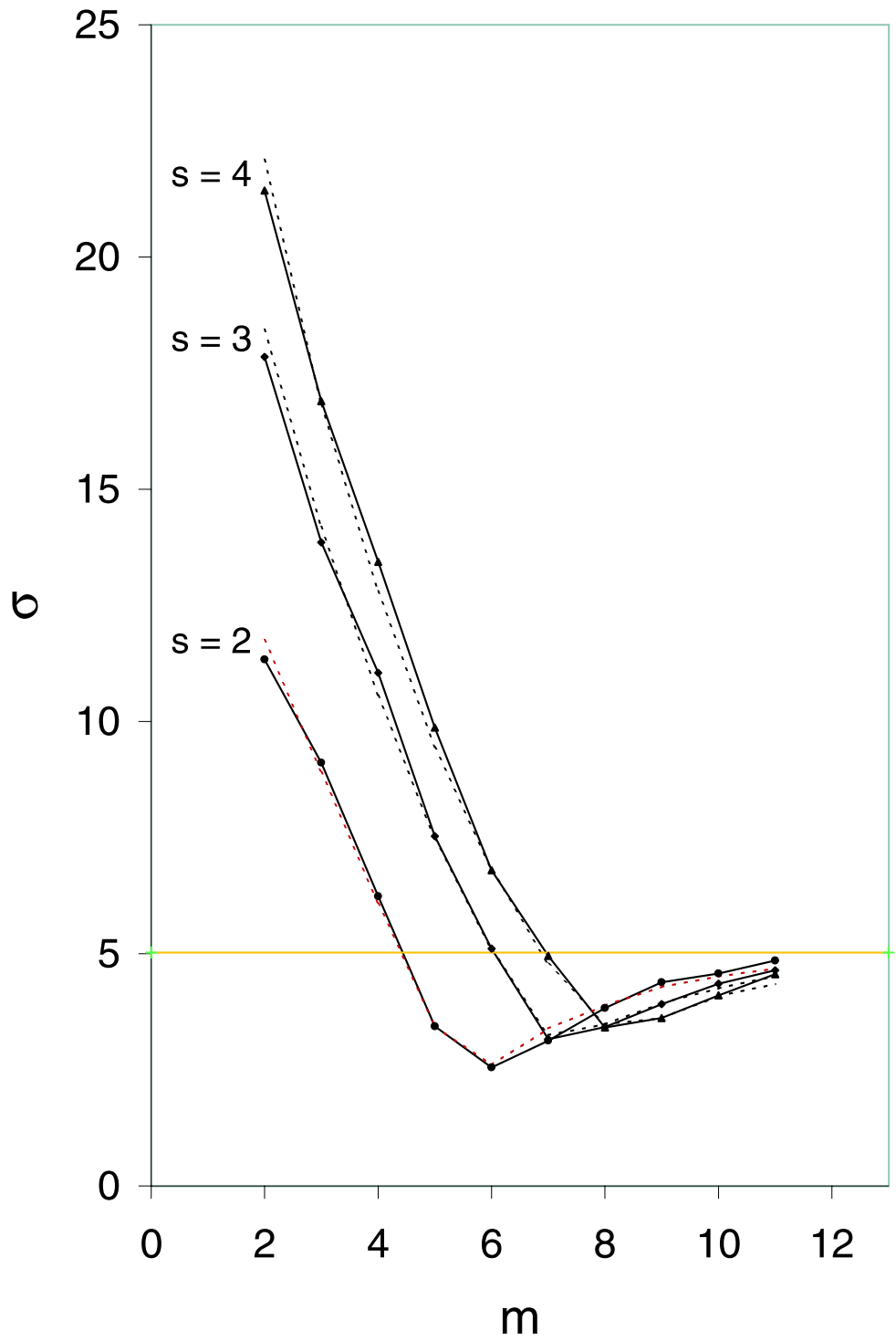
\noindent FIGURE 7. Proof-of-concept of Crowd-Anticrowd approach. Graph shows
volatility for the Minority Game as a function of memory size $m$, for
$s=2,3,4$ strategies per agent, and $N=101$ agents. Solid curve: numerical
simulation. Dashed curve: Crowd-Anticrowd theory, evaluated numerically using
Equation 13. Random (coin-toss) limit $\protect\sigma =\protect\sqrt{N}/2=5.0$ is
indicated.

\vskip0.1in
\noindent FIGURE 8 Schematic representation of how the scores of three strategies
and their three anti-correlated partners, might vary in time.

\vskip0.1in
\noindent FIGURE 9 Same as Figure 8, but with strategies now ranked  in terms of
virtual-point ranking
$K$. Hence shows the
variation of the scores of the two top scoring strategies from Figure 8. 

\vskip0.1in
\noindent FIGURE 10  Schematic representation of 
$\Omega $ with $m=2$, $s=2$, in the RSS. The strategies are ranked according to
the value of $K$. The shaded elements represent those agents whose highest scoring
strategy is ranked 4'th highest in score, i.e. $K=4$.

\vskip0.1in
\noindent FIGURE 11 Analytic forms of volatilities $\sigma _{1}^{delta\text{ }f}$\ ,
$\sigma _{1}^{flat\text{  }f}$\ and$\ \sigma _{1}^{flat\text{ }f,\text{ }high\text{
}m}$, as a function of $m$ for $s=2$, $4$ and $8$. Also shown are the numerical
values obtained from different simulation runs (triangles, crosses and circles). The
results for
$s=2$ are the same as Figure 2.

\vskip0.1in
\noindent FIGURE 12 Comparison between the theoretical values of $n_{Q}$ calculated
using Equation \ref{YofR} for $s=2$ and $N=101$ (dashed lines) and the numerical
values taken from the MG simulation (solid lines). 

\vskip0.1in
\noindent FIGURE 13 Form of $f_{Q^{\prime },\overline{Q}}$ for $Q=1$ as a function
of $Q^{\prime }$, taken from the numerical MG simulation at $m=2,5$ and $10$.

\vskip0.1in
\noindent FIGURE 14 Volatility as a function of memory $m$ for $s=2$ and $N=101$. 
The spread of numerical values for different runs of the MG (open circles) is
compared to the Crowd-Anticrowd theoretical calculation using Equation
\ref{fq} with numerically obtained values of $f_{Q^{\prime },\overline{Q}}$ (solid
circles).

\vskip0.1in
\noindent FIGURE 15 Analytic theory for the volatility curves
(solid lines: see Section 4.3) for $s=2$, $s=4$ and $s=6$ with
$N=101$. Numerical simulations (dashed line) also shown, averaged over a limited
number of runs (hence the jaggedness).


\begin{thebibliography}{99}
\bibitem{general} Note that a single $\Omega$ `macrostate' can still correspond to
many possible `microstates' for partitions of strategy allocation among the agents.
Hence the Crowd-Anticrowd theory retained at the level of a given specified
$\Omega$, will still be of use for all games whose initital strategy allocations
correspond to that same $\Omega$.
\end{thebibliography}
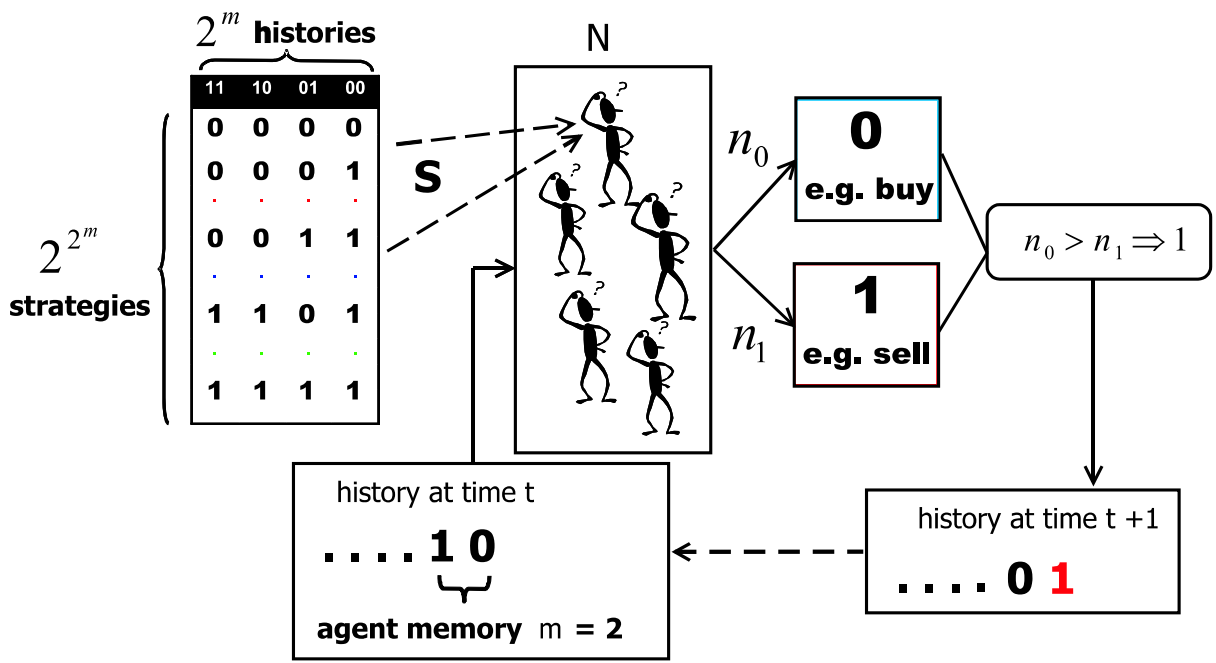
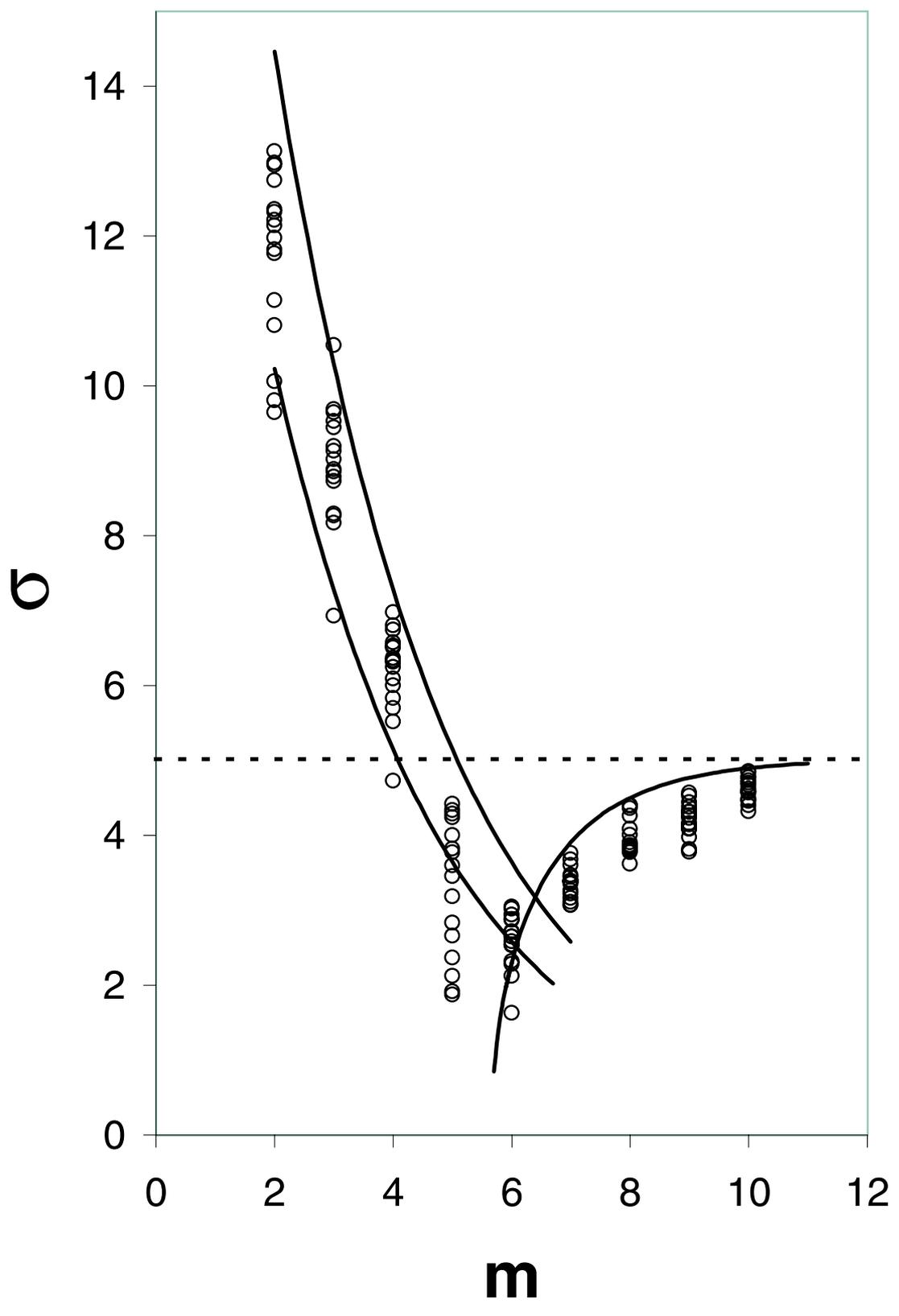
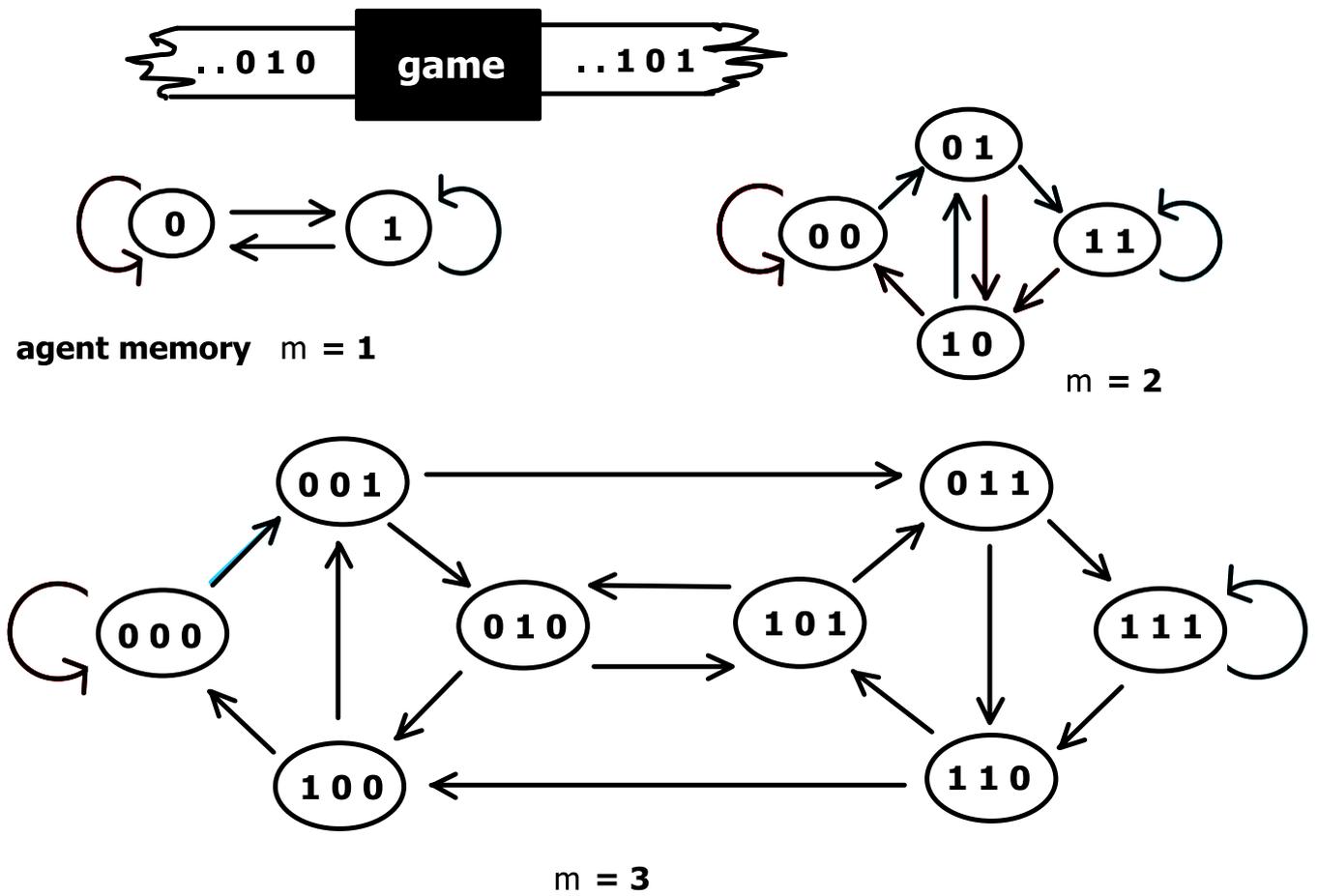
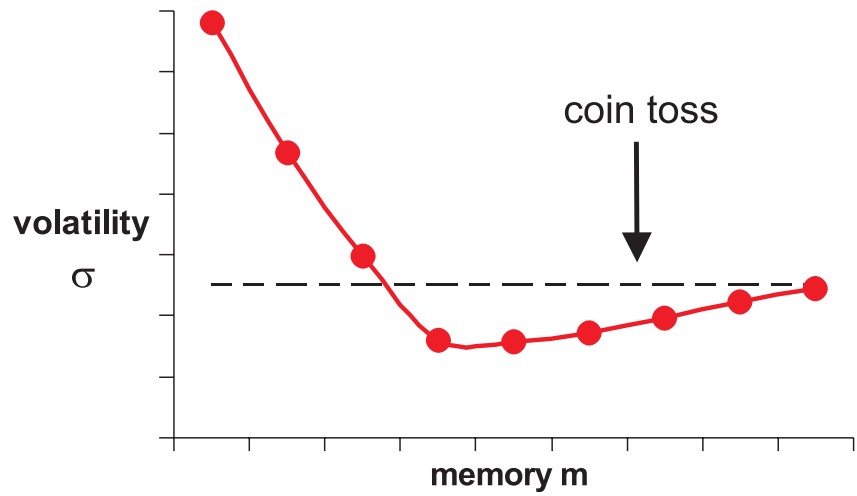
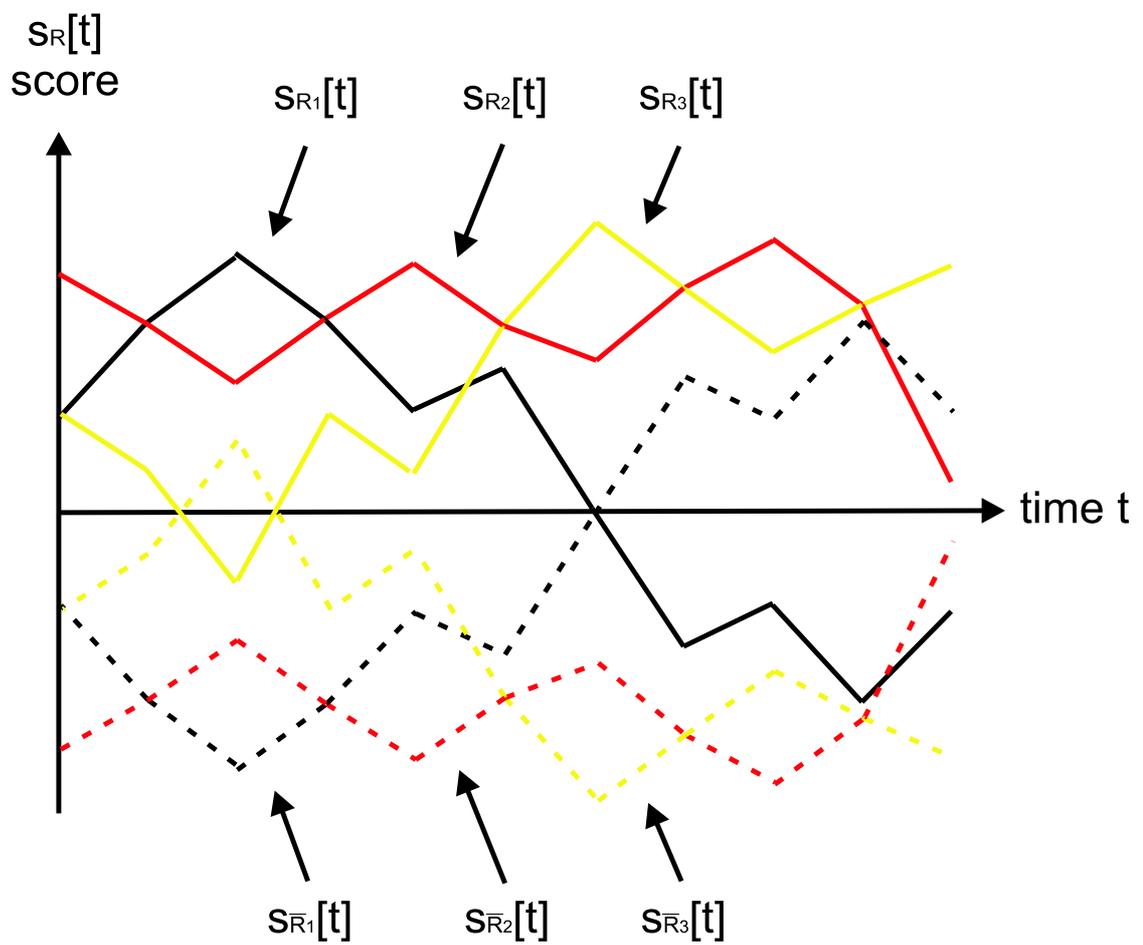
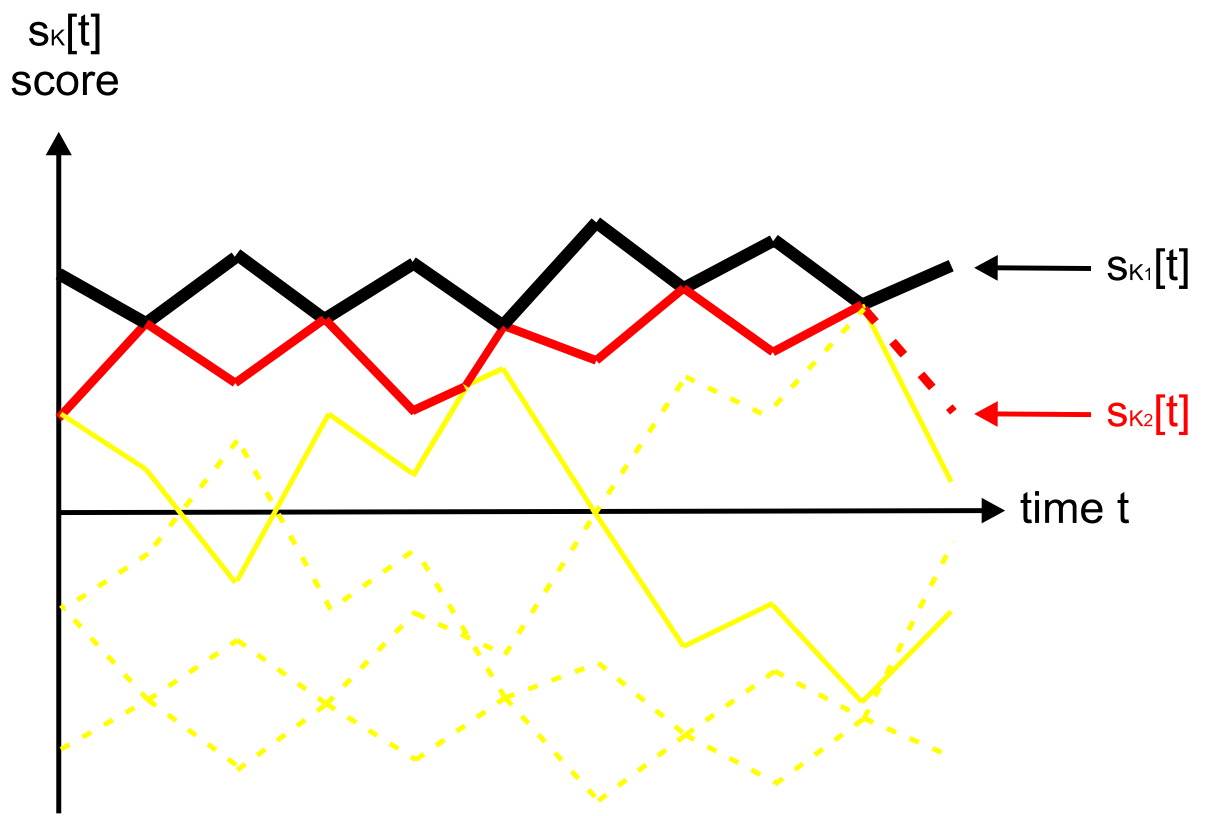
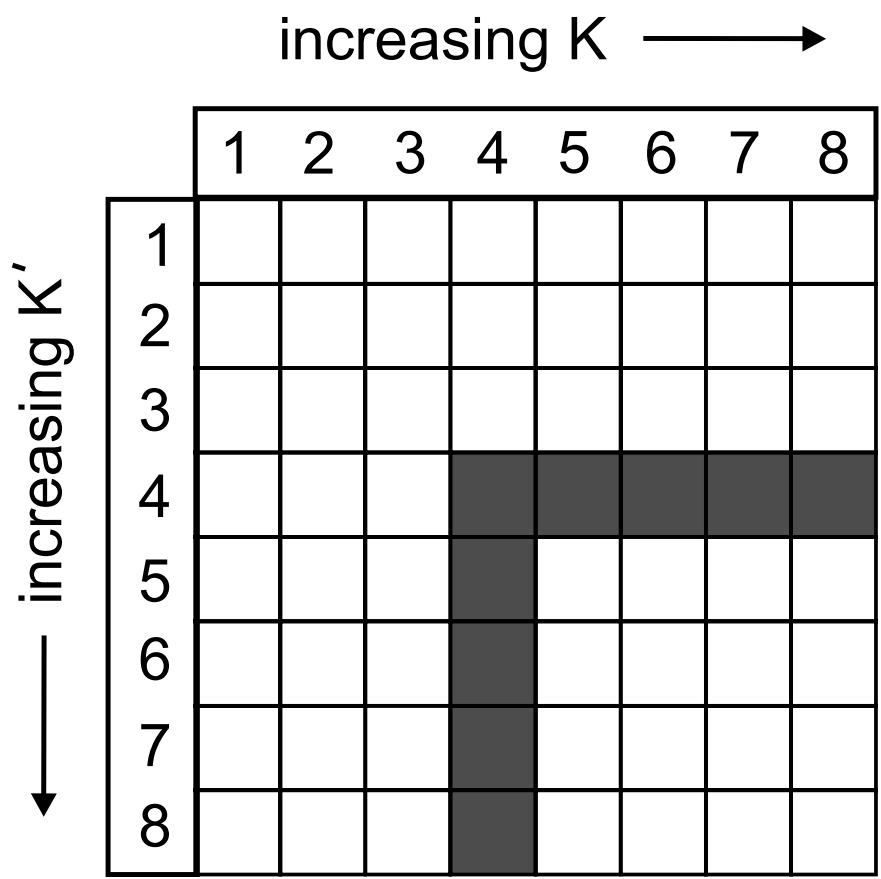
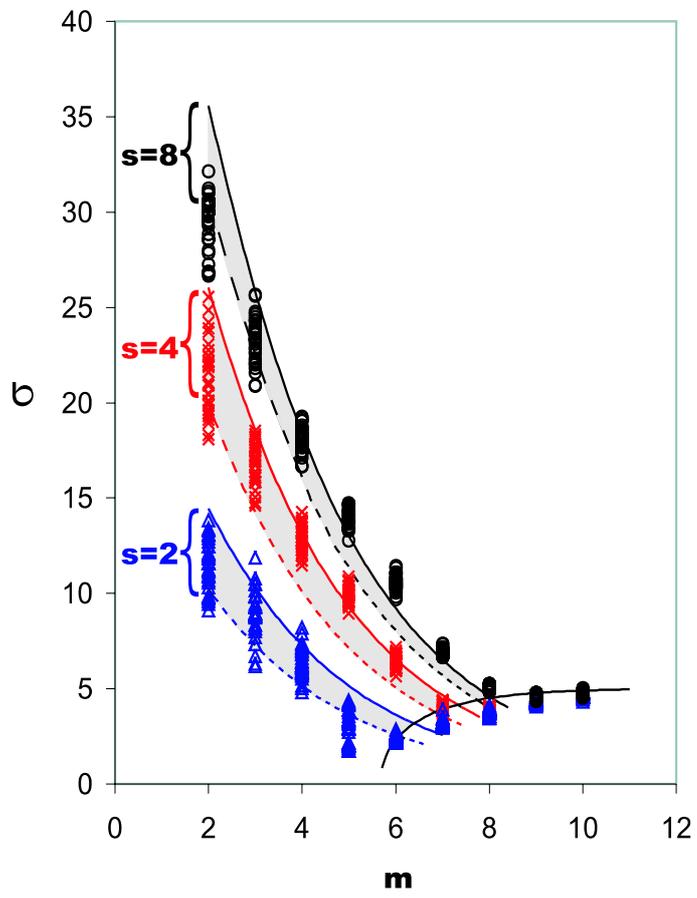
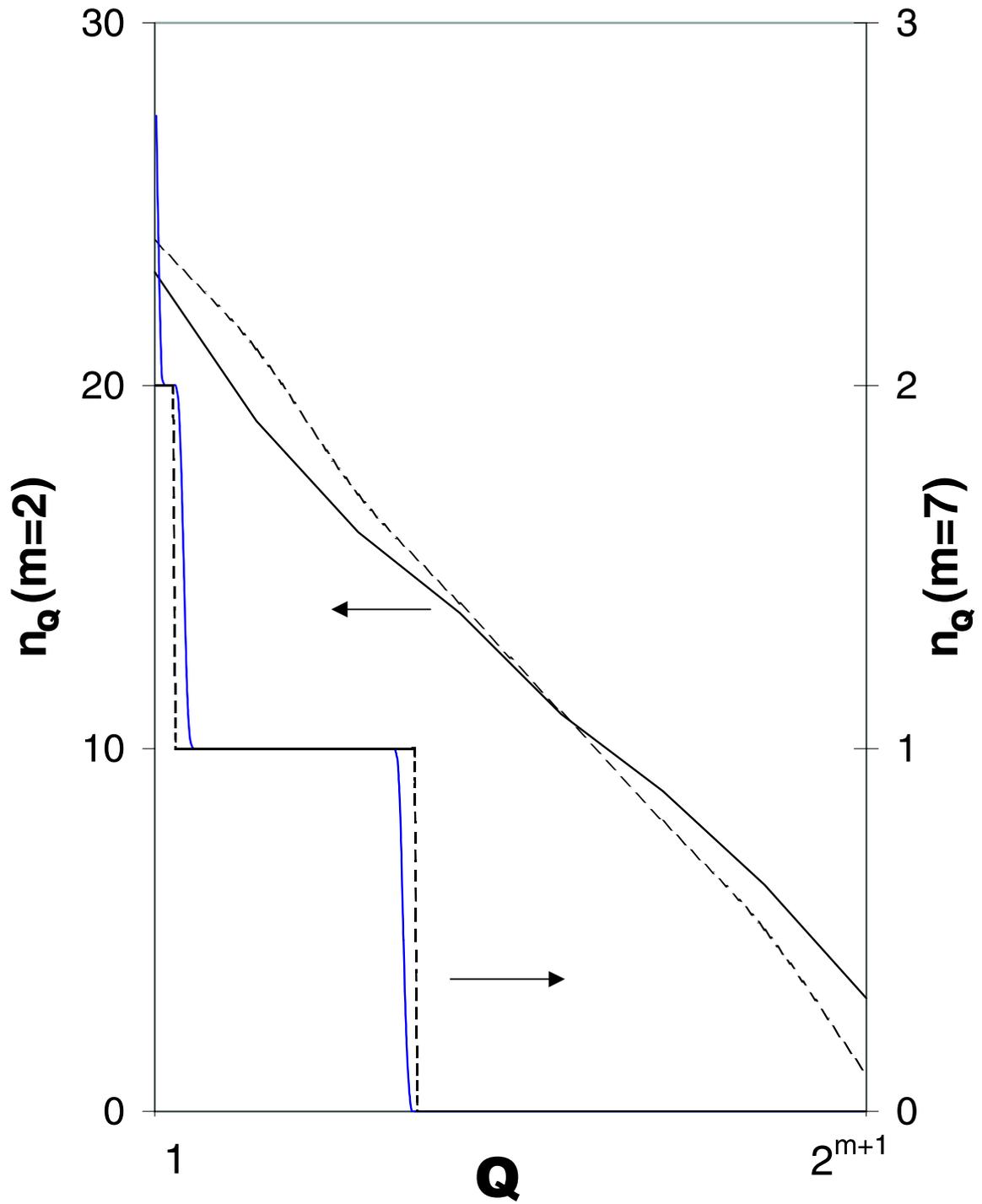
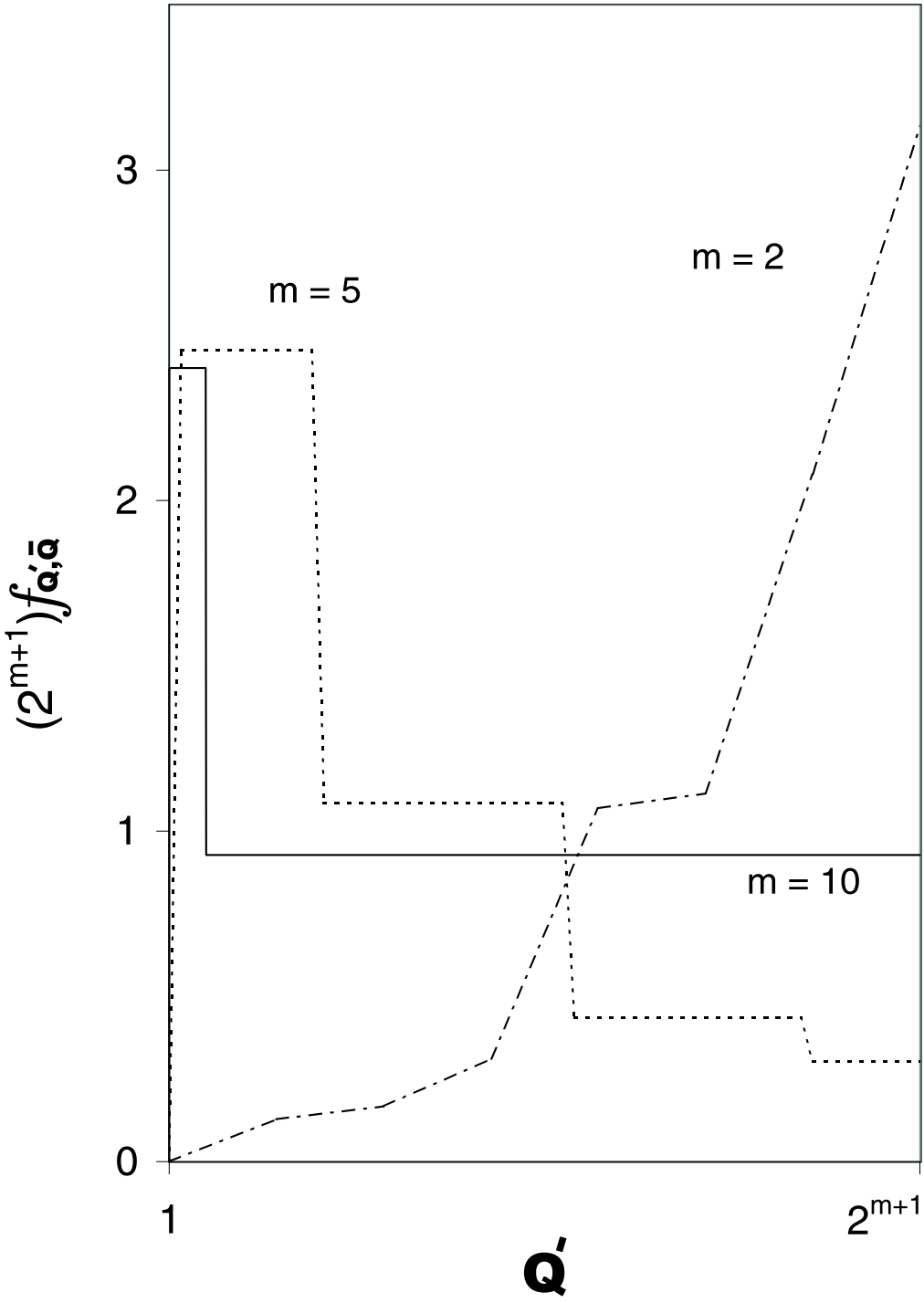
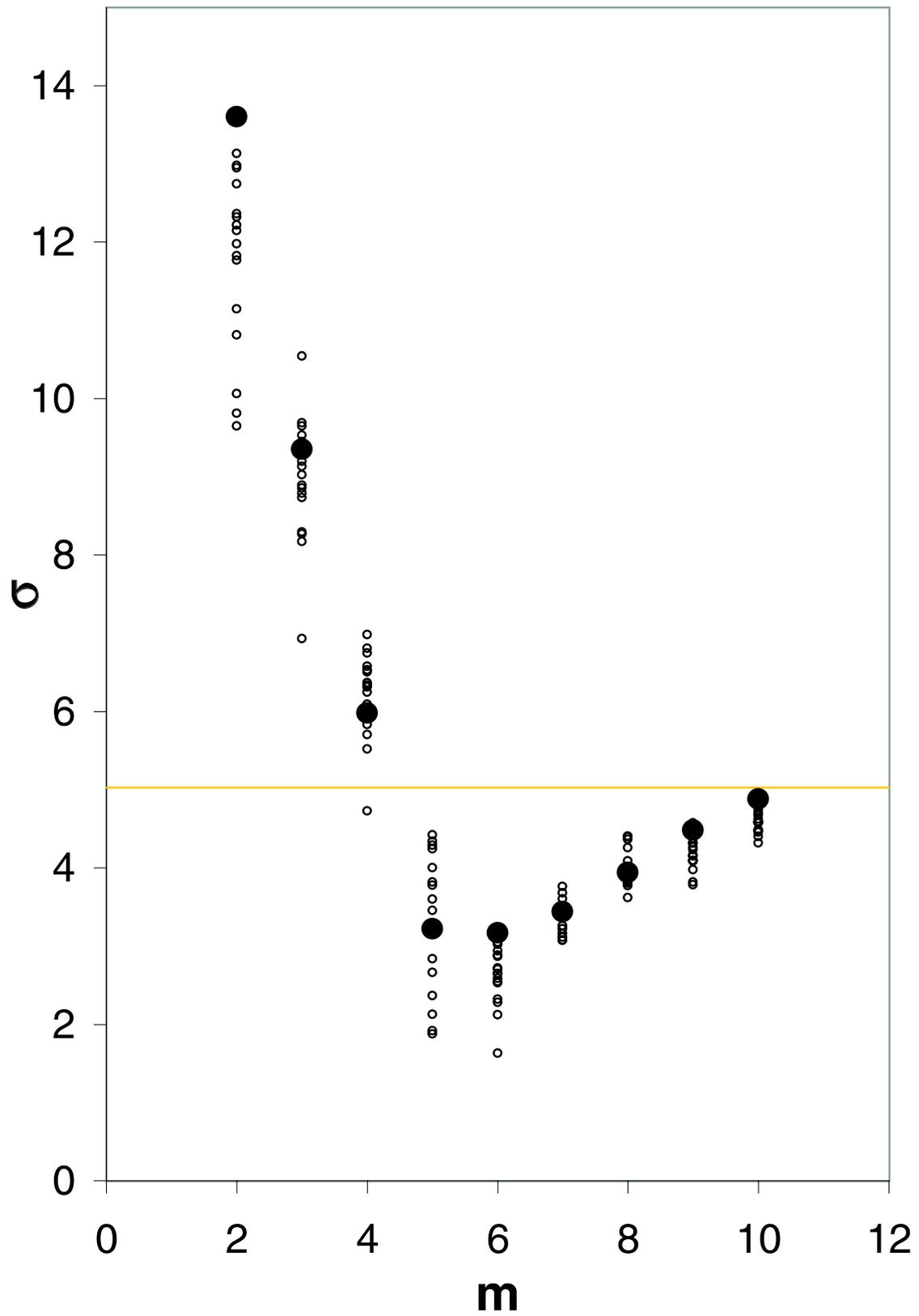
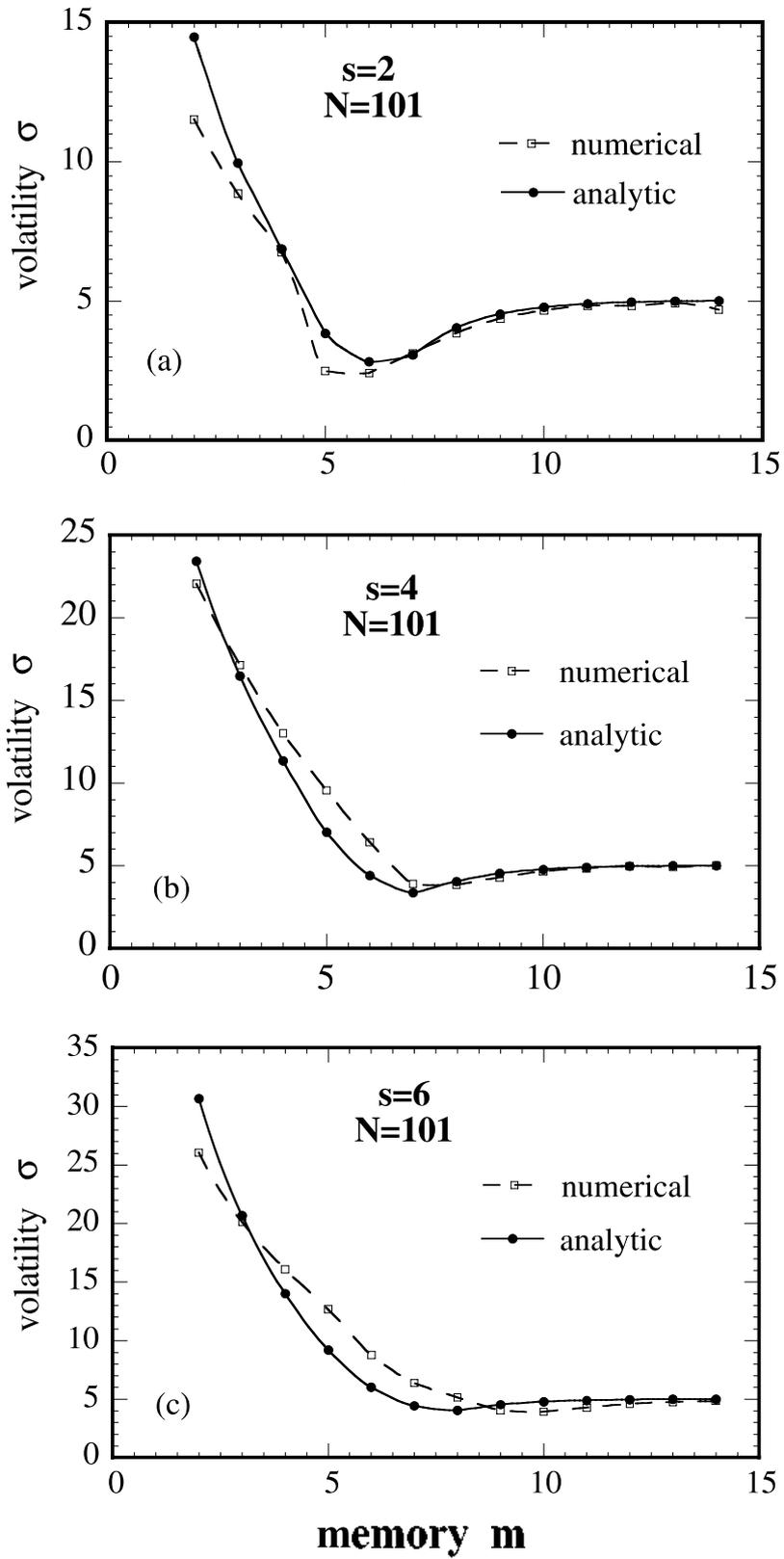
\end{document}